\documentclass[twocolumn, superscriptaddress,amsfont,amssymb,amsmath, showpacs,balancelastpage, nofootinbib]{revtex4-2}
\usepackage{graphicx,longtable,mathrsfs,color,array}
\usepackage[unicode=true,pdfusetitle,
bookmarks=true,bookmarksnumbered=true,bookmarksopen=true,bookmarksopenlevel=1,
breaklinks=false,pdfborder={0 0 0},backref=false,colorlinks=true]{hyperref}
\hypersetup{citecolor=blue,filecolor=blue,linkcolor=blue,urlcolor=blue}
\usepackage[usenames,dvipsnames]{xcolor}
\usepackage{amssymb,amsmath,mathtools,mathrsfs,enumitem}
\usepackage{epsfig,subfigure,placeins,float}
\usepackage{booktabs,multirow}
\usepackage{exscale,relsize}
\usepackage[normalem]{ulem}
\usepackage[T1]{fontenc}
\usepackage[utf8]{inputenc}
\usepackage{enumerate}
\usepackage{times, mathptmx}
\usepackage{tikz}
\usepackage{aas_macros}
\usetikzlibrary{arrows,positioning,decorations.markings,decorations.pathmorphing,calc}

\usepackage[capitalise]{cleveref}
\usepackage{diagbox}
\usepackage{hhline}

\newcommand{\be}{\begin{equation}}
\newcommand{\ee}{\end{equation}}

\newcommand{\comment}[1]{}

\newcommand{\mSt}{\{\mu_{\rm{today}},\Sigma_{\rm{today}}\}}
\newcommand{\mt}{\mu_{\rm{today}}}
\newcommand{\St}{\Sigma_{\rm{today}}}
\newcommand{\BMz}{\{\alpha_{B,0},\alpha_{M,0}\}}
\newcommand{\wzwa}{\{w_0, w_a\}}
\newcolumntype{C}[1]{>{\centering\let\newline\\\arraybackslash\hspace{0pt}}m{#1}}

\newcommand{\togglered}[1]{#1}

\definecolor{hyperref}{RGB}{026,028,087}
\def\gsim{ \lower .75ex \hbox{$\sim$} \llap{\raise .27ex \hbox{$>$}} }
\def\lsim{ \lower .75ex \hbox{$\sim$} \llap{\raise .27ex \hbox{$<$}} }

\graphicspath{{./figures/}}
\allowdisplaybreaks

\tikzstyle{vecArrow} = [thick, decoration={markings,mark=at position
1 with {\arrow[semithick]{open triangle 60}}},
double distance=1.4pt, shorten >= 5.5pt,
preaction = {decorate},
postaction = {draw,line width=1.4pt, white,shorten >= 4.5pt}]

\begin{document}
\title{Dark energy constraints in light of theoretical priors}

\author{Neel Shah}
\affiliation{Institute of Cosmology \& Gravitation, University of Portsmouth, Portsmouth, PO1 3FX, U.K.}
\affiliation{Department of Physics \& Astronomy, University College London, London, WC1E 6BT, U.K}

\author{Kazuya Koyama}
\affiliation{Institute of Cosmology \& Gravitation, University of Portsmouth, Portsmouth, PO1 3FX, U.K.}

\author{Johannes Noller}
\affiliation{Department of Physics \& Astronomy, University College London, London, WC1E 6BT, U.K}
\affiliation{Institute of Cosmology \& Gravitation, University of Portsmouth, Portsmouth, PO1 3FX, U.K.}

\begin{abstract}
In order to derive model-independent observational bounds on dark energy/modified gravity theories, a typical approach is to constrain parametrised models intended to capture the space of dark energy theories. 
Here we investigate in detail the effect that the nature of these parametrisations can have, finding significant effects on the resulting cosmological dark energy constraints. 
In order to observationally distinguish well-motivated and physical parametrisations from unphysical ones, it is crucial to understand the theoretical priors that physical parametrisations place on the phenomenology of dark energy.
To this end we discuss a range of theoretical priors that can be imposed on general dark energy parametrisations, and their effect on the constraints on the phenomenology of dynamical dark energy. More specifically, we investigate both the phenomenological $\{\mu,\Sigma\}$ parametrisation 
as well as effective field theory (EFT) inspired approaches to model dark energy interactions. We compare the constraints obtained in both approaches for different phenomenological and theory-informed time dependences for the underlying functional degrees of freedom, discuss the effects of priors derived from gravitational wave physics, and investigate the interplay between constraints on parameters constraining only the background evolution vs. parameters controlling linear perturbations.
\end{abstract}

\date{\today}
\maketitle

\tableofcontents

\section{Introduction} \label{sec-intro}

When probing the nature of dark energy, especially in the context of establishing whether dark energy is dynamical or not, a choice has to be made how to parametrise the freedom associated with this dynamics (even if the choice is to perform a `non-parametric', e.g. binned, fit). Current surveys -- see e.g. \cite{Planck:2018vyg,DES:2022ccp,Ishak:2024jhs} -- and indeed the field at large, therefore tend to adopt a two-pronged approach when faced with this choice, adopting I) a phenomenological parametrisation of this freedom with minimal underlying theoretical assumptions, and II) a complementary more theory-informed approach, which imposes additional theory priors that capture large classes of well-defined theory space. In this paper we investigate the effect of different types of theoretical priors on constraints on the nature of dark energy, focusing in particular on the physics of linear perturbations around cosmological backgrounds which yields some of the (if not the) strongest bounds in this context. 
\\

Cosmological scalar metric perturbations at the linear level can be described by two metric potentials:
\begin{gather}
ds^2 = -(1 + 2\Phi)dt^2 + a^2(1 - 2\Psi)dx_idx^i\ ,
\end{gather}
where we have written the metric in the Newtonian gauge, and have assumed a spatially flat FLRW background motivated by strong observational constraints on spatial curvature \cite{Planck:2018vyg}.

In the simplest dark energy models, where the role of dark energy is played by a cosmological constant $\Lambda$ or by a canonical scalar field as in quintesssence \cite{Ratra:1987rm,Wetterich:1987fm,Frieman:1995pm,Coble:1996te,Caldwell:1997ii}, the two metric potentials $\Phi$ and $\Psi$ are equal in our regime of interest (namely the late universe, when only dark matter and dark energy contribute appreciably to the evolution of perturbations). However, in more general models of dark energy, for instance ones that involve a scalar field with a nontrivial coupling to gravity (as can readily be found within the Horndeski class of scalar-tensor theories \cite{Horndeski:1974wa,Charmousis:2011bf,Deffayet:2011gz,Kobayashi:2011nu}), they can evolve differently. 
At a phenomenological level, this can be captured by modifying two versions of the Poisson equation:
\begin{gather}\label{Poisson_mu_scaledep}
\togglered{\dfrac{k^2}{a^2}\Phi = -4\pi G\mu(a, k)\rho\Delta\ ,}\\ \label{Poisson_Sigma_scaledep}
\togglered{\dfrac{k^2}{a^2}(\Phi+\Psi) = -8\pi G\Sigma(a, k)\rho\Delta\ ,}
\end{gather}
where G is Newton's constant, and $\Delta$ is the density contrast in the comoving gauge. The form of these Poisson equations is motivated by the fact that the gravitational potentials appearing in the left hand sides of \cref{Poisson_mu} and \cref{Poisson_Sigma} are the same potentials that appear in the geodesic equations for nonrelativistic and relativistic matter respectively. This means that the evolution of $\Phi$ can be probed by gravitational clustering, while that of $\Phi + \Psi$ can be probed by gravitational lensing. A deviation of $\mu(a, k)$ or $\Sigma(a, k)$ from unity at any redshift \togglered{or scale} indicates that dark energy affects structure formation directly through the evolution of metric perturbations, over and above the indirect effect from a background expansion history potentially different from one given by a cosmological constant.

A commonly used theory-informed approach to parametrising dark energy dynamics is the so-called Effective Field Theory of Dark Energy (EFTDE) \cite{Gubitosi:2012hu, Bloomfield:2012ff}. This approach identifies the degree of freedom (the Goldstone boson) associated with the breaking of time translation invariance -- a breaking that naturally happens in cosmology -- and systematically constructs the possible interactions of this degree of freedom compatible with the (residual) symmetries of FLRW backgrounds. 
Note closely related approaches, where diffeomorphism invariance is imposed from the start and general interactions consistent with an FLRW background are constructed for general field content (e.g. a scalar, vector or tensor) \cite{Lagos:2016wyv,Lagos:2017hdr}.
When restricting to theories with an effective additional scalar degree of freedom (effectively the Goldstone boson identified above) and up to second order derivatives in the equations of motion (as we will do in this paper), the resulting descriptions of linear cosmological perturbations are equivalent to working with the Horndeski class of scalar-tensor gravity models linearised on a cosmological background \cite{Kobayashi:2011nu,Bellini:2014fua}.

In the EFTDE/Horndeski framework, the linearised dynamics of cosmological perturbations are completely described by five functions of time \cite{Bellini:2014fua, Gleyzes:2014rba}. These are: 1) $H(a)$, or equivalently, the DE equation of state $w(a)$ characterising the background expansion history, 2) $\alpha_K(a)$, called the kineticity, which characterises the pure kinetic energy of the scalar field independent of kinetic mixing with the metric, 
3) $\alpha_B(a)$, called the braiding, which characterises kinetic mixing between the scalar field and the metric, 4) $M^2(a)$ or its rate of change $\alpha_M(a) \equiv d \ln(M^2)/d \ln(a)$, with $M^2(a)$ being the effective Planck mass seen by tensor perturbations, and 5) $\alpha_T(a)$ characterising the speed of gravitational waves (which can be different from the speed of light, always set to unity). These functions completely determine the phenomenology of linear perturbations. \togglered{The Poisson equations in these theories reduce to \cref{Poisson_mu_scaledep,Poisson_Sigma_scaledep} under the quasistatic approximation (QSA, described in detail in \cref{sec-mapping}). Moreover, when the mass of the scalar field is small, one can also neglect the scale-dependence of $\{\mu, \Sigma\}$, and the Poisson equations in such theories becomes}

\begin{gather}\label{Poisson_mu}
\togglered{\dfrac{k^2}{a^2}\Phi = -4\pi G\mu(a)\rho\Delta\ ,}\\ \label{Poisson_Sigma}
\togglered{\dfrac{k^2}{a^2}(\Phi+\Psi) = -8\pi G\Sigma(a)\rho\Delta\ ,}
\end{gather}
\togglered{where one can derive the $\{\mu(a), \Sigma(a)\}$ functions from a given set of $\{\alpha_i(a), w(a)\}$. In the following, we will work with these scale-independent Poisson equations for simplicity, with the QSA and the scale-independence of the Poisson equations justified in \cref{sec-mapping}. We will also assume scale-independence of $\{\mu, \Sigma\}$ in the Poisson equations in the purely phenomenological approach in order to enable an apples-to-apples comparison with the theory-informed approach, where the $\{\mu, \Sigma\}$ derived from the EFTDE models we consider are in fact scale-independent.}

In both the above approaches, the main practical challenge for observational constraints is that we cannot constrain arbitrary functions of redshift with a finite amount of data. In order for the data to have non-negligible constraining power, typically the time dependence of $\{\mu, \Sigma\}$ or $\{\alpha_i\}$ is heavily restricted via parametric forms involving only one or two parameters, and it is these one or two parameters that are then constrained with the data. Without any theoretical input on such parametrisations, it is easy to miss well-motivated theories that could be observationally distinguished from GR, or to falsely interpret phenomenological signs of deviations from GR that might be impossible to obtain from an underlying physical theory. This is why it is very useful to impose theoretical priors on the dynamical DE functions. For instance, when considering $\{\mu, \Sigma\}$, the constraint of deriving them within the EFTDE framework (i.e. from underlying $\{\alpha_i\}$) amounts to a strong theoretical prior which imposes a correlation between their present day values. One can also go a step further and impose theoretical priors on the $\{\alpha_i\}$. In this paper, we investigate how imposing and relaxing various theoretical priors on the $\{\alpha_i\}$ affects constraints on the phenomenology of DE perturbations, i.e. on $\{\mu, \Sigma\}$. The priors we consider come in various shapes and forms: on the functional form of the $\{\alpha_i\}$, which $\{\alpha_i\}$ to consider in the first place etc., from underlying theoretical considerations relating to large scale structure cosmology, gravitational waves, and the Lagrangian of the EFTDE.

{\bf Outline}: The paper is organised as follows. In \cref{sec-Param} we explain the phenomenological and EFTDE approaches that we use to parametrise DE dynamics. In \cref{sec-mapping}, we connect the theory-informed and phenomenological approaches by deriving the theoretical prior from the EFTDE on the present-day values of phenomenological modifications to gravity, and discuss various features of the mapping from the EFTDE parameters to the phenomenological modified gravity parameters. In \cref{sec-data}, we describe the datasets that we use to obtain observational constraints, and our methodology of obtaining the constraints using an MCMC analysis. In \cref{sec-phenoEFT}, we compare observational constraints on modified gravity phenomenology between the EFTDE and phenomenological parametrisations, and explain the difference in the constraints using the EFTDE prior on the phenomenology, and the difference in time evolutions of the phenomenological modifications to gravity across the two parametrisations. In \cref{sec-scale}, we look at the effect of assuming a different time dependence of the functional degrees of freedom in the EFTDE and phenomenological parametrisations, and explain qualitative differences in how the EFTDE parameters map to phenomenological modified gravity parameters across the two time dependences In \cref{sec-shiftsym}, we look at a theoretically motivated time dependence for the functional degree of freedom in a restricted class of EFTDE theories, and check whether it can be distinguished from naive phenomenological parametrisations at the level of constraints on the present day phenomenology. In \cref{sec-GWpriors}, we look at two theoretical priors on the EFTDE motivated by gravitational wave physics: I) a prior on the speed of GWs on cosmological scales, and II) a prior that prevents gradient instabilities of DE perturbations in the presence of a background of GWs, such as the one sourced by colliding black holes. In \cref{sec-background}, we look at the interplay between the constraints on the background and perturbation dynamics in various phenomenological and EFTDE parametrisations. Finally, we present our conclusions and an outlook in \cref{sec-conclusions}.

\section{Parametrising Dark Energy} \label{sec-Param}

Dark energy affects the background expansion of the universe as well as the evolution of perturbations. In model-agnostic approaches, these two effects are constrained independently. 
While this decoupling of background and perturbative dynamics does not take place in setups such as Quintessence or k-essence \cite{Caldwell:2005tm,Armendariz-Picon:2000nqq}, in general scalar-tensor theories (e.g. full Horndeski) there is sufficient freedom for the background expansion and behaviour of perturbations to indeed be decoupled in this way. There are therefore large classes of known theories for which this ansatz is consistent, serving as motivation for its use in model-agnostic setups as well.

It is useful to emphasise upfront that there are at least two ways in which such model-agnostic approaches can be theory-informed (while still remaining non-theory-specific): I) by informing what the underlying functions controlling the evolution are in the first place (e.g. which $\{\alpha_i\}$ are non-zero and independent), and II) by specifying what functional form these functions take. We will use the word "parametrisation" to refer to both cases: i.e. both for specifying the functional degrees of freedom that control the evolution of perturbations, and for specifying the time evolution of those free functions.

\subsection{Parametrising the background}

At the level of the background expansion, the relevant freedom is encoded in a single function of scale factor (or, equivalently, of time or of redshift). This is frequently parametrised not via the expansion rate $H(a)$, which (as a relatively directly measurable quantity) is closer to the phenomenology, but rather via the DE equation of state $w(a)$. Of course, once either $H(a)$ or $w(a)$ is given, the other can be derived via the Friedmann equations. 

Arguably the most common parametrisation used for $w(a)$ is the  Chevallier-Polarski-Linder (CPL) parametrisation \cite{Chevallier:2000qy,Linder:2002et}, effectively a lowest order Taylor expansion in the scale factor:
\begin{gather}\label{w0wa}
w(a) = w_0 + w_a(1-a)\ .
\end{gather}

One may wonder how restrictive such a parametrisation is. After all, suppressed higher order terms in the Taylor expansion in principle have sizeable effects at relevant redshifts -- for example, one can naively estimate that a quadratic $w_{a^2}(1-a)^2$ term in the equation of state parametrisation will yield ${\cal O}(5\%)$ corrections at redshift $z_{\rm eq} \sim 0.3$, where dark energy begins to dominate.\footnote{For this estimate, we are assuming $w_0,w_a, w_{a^2}, \ldots$ coefficients are ${\cal O}(1)$.} Ignoring such higher order corrections then potentially seems at odds with the aim of predicting observables to percent level accuracy, as is common for upcoming surveys \cite{DESI:2023dwi,Euclid:2024yrr,LSSTDarkEnergyScience:2018jkl}.
However, somewhat surprisingly, the parametrisation \eqref{w0wa} can do better than naively expected in a very concrete sense: Specifically, both in the case of thawing quintessence as well as for shift-symmetric Horndeski models with luminally propagating gravitational waves, it was found that a $\{w_0,w_a\}$ fit to measured observables can indeed reproduce these with the accuracy required \cite{Garcia-Garcia:2019cvr,Traykova:2021hbr}.\footnote{Here it is worth emphasising that this does not mean the equation of state is reconstructed with percent level accuracy itself, but that observables such as the Hubble rate, the angular diameter distance, and the growth factor are reproduced with the required accuracy.} While this is not a universally true statement, e.g. the above is not found to be true for tracking quintessence \cite{Garcia-Garcia:2019cvr}, it provides further motivation for the use of the CPL parametrisation over and above its common use. 

In this paper, our primary focus is on constraining DE perturbations, and our main interest in constraining the background is only for studying the impact it has on the perturbations constraints. For this reason, we fix a $\Lambda{}$CDM background with $w(a)=-1$ in most of our analysis, focusing on studying the effects of various theoretical priors on the dynamics of perturbations on top of this background as is commonly done (for example, in \cite{Bellini:2015xja,Kreisch:2017uet,Noller:2018wyv,Perenon:2019dpc,SpurioMancini:2019rxy,Arjona:2019rfn,Melville:2019wyy,Arjona:2019rfn,Andrade:2023pws,Seraille:2024beb}). Complementary to this, we study the effect of freeing up the background in \cref{sec-background}, empirically justifying the previous approach (fixing the background and solely studying changes to the perturbative dynamics) in the process, while also pointing out an interesting exception where the background and perturbation dynamics do not decouple.

\subsection{$\mu$ and $\Sigma$: Phenomenological parametrisations}\label{subsec-Sigmu}

Like for $w(a)$, there is a choice to be made about how to parametrise the time dependence of $\{\mu(a), \Sigma(a)\}$ introduced in \cref{Poisson_mu,Poisson_Sigma}. The only physical motivation for this choice is the expectation that if the same underlying theory determines both the DE background and perturbations, then DE effects on perturbations should become relevant at the same time as the effects on the background. Thus, deviations from GR are assumed to be proportional to the fractional energy density of dark energy $\Omega_{DE}$:
\begin{gather}
\mu(a) - 1 = (\mu_0 - 1)\dfrac{\Omega_{DE}(a)}{\Omega_{DE,0}}\ ,\\ \label{Sigma_OmegaDE}
\Sigma(a) - 1 = (\Sigma_0 - 1)\dfrac{\Omega_{DE}(a)}{\Omega_{DE,0}}\ .
\end{gather}
This is the parametrisation used by several recent LSS surveys \cite{DES:2022ccp, Ishak:2024jhs}\footnote{Note that our definitions of $\{\mu(z), \Sigma(z)\}$ and $\{\mu_0, \Sigma_0\}$ differ slightly from those of \cite{DES:2022ccp, Ishak:2024jhs}.}. Closely related parametrisations explicitly work with the $\gamma(a)$ function instead, which satisfies\footnote{This relation between $\gamma$ and $\{\mu,\Sigma\}$ neglects any difference between the two potentials coming from anisotropic stress in the stress-energy tensor of matter. This is a valid approximation in the late universe with standard CDM.}
\begin{gather}
\gamma(a) \equiv \dfrac{\Psi}{\Phi} = 2\dfrac{\Sigma(a)}{\mu(a)} - 1.
\end{gather}
and therefore directly quantifies the difference between the two gravitational potentials $\Phi$ and $\Psi$.
Clearly specifying any two out of the three functions $\{\mu,\Sigma,\gamma\}$ also determines the third, i.e. only two of those three functions are independent. Instead of parametrising $\{\mu,\Sigma\}$ and deriving $\gamma$, one can therefore instead e.g. specify $\{\mu,\gamma\}$ and derive $\Sigma$. This approach is e.g. taken by the Planck collaboration \cite{Planck:2018vyg}, where these functions are parametrised (following earlier work such as \cite{Simpson:2012ra}) as
\begin{gather}\label{mu_OmegaDE}
\mu(a) - 1 = (\mu_0 - 1)\dfrac{\Omega_{DE}(a)}{\Omega_{DE,0}}\ ,\\ \label{gamma_OmegaDE}
\gamma(a) - 1 = (\gamma_0 - 1)\dfrac{\Omega_{DE}(a)}{\Omega_{DE,0}}\ .
\end{gather}
In our analysis, when considering phenomenological parametrisations, we will parametrise $\{\mu,\gamma\}$. We show in \cref{app-DESode} that the observational posteriors on the present day values of $\mu$ and  $\Sigma$ obtained with this parametrisation are very similar to those from parametrising $\{\mu, \Sigma\}$.

\togglered{While the time dependence in \cref{mu_OmegaDE,gamma_OmegaDE} is the fiducial one that we assume in most of our analyses, in \cref{sec-scale} we compare it with another phenomenological time dependence where these functions are proportional to the scale factor. Here, they are given by}
\begin{gather}\label{mu_scale}
\togglered{\mu(a) - 1 = (\mu_0 - 1)a\ ,}\\ \label{gamma_scale}
\togglered{\gamma(a) - 1 = (\gamma_0 - 1)a\ .}
\end{gather}

For comparing different modified gravity parametrisations (either phenomenological or theory-informed), it is useful to compare constraints on a quantity with a parametrisation-independent definition that can be derived in any parametrisation. For this purpose, we use the present day values of $\{\mu, \Sigma\}$, denoted $\mSt$:\footnote{We avoid the notation $\mu_0, \Sigma_0$ for present day values because these symbols are frequently used in the definitions of phenomenological parametrisations like \cref{mu_OmegaDE,gamma_OmegaDE} but seldom in the context of theory-informed parametrisations such as those in \cref{subsec-eftde}. We want to emphasise that it makes sense to define the present day values of $\mu, \Sigma$ using \cref{mu_today,Sigma_today} without any reference to a particular phenomenological or theory-informed parametrisation of modified gravity.}
\begin{gather}\label{mu_today}
\mt \equiv \mu(a = 1)\ ,\\ \label{Sigma_today}
\St \equiv \Sigma(a = 1)\ .
\end{gather}

For example, for both parametrisations \cref{mu_OmegaDE,gamma_OmegaDE} and \cref{mu_scale,gamma_scale}, the present day values are given by $\mt = \mu_0$, and $\St = \tfrac{1}{2}\mu_0(1 + \gamma_0)$. An informative classification of the entire space of $\mt,\St$ values is into four quadrants around the GR limit: $\mt < 1 / > 1, \St < 1 / > 1$. 

This classification is very instructive in the context of the theory-informed parametrisations of \cref{subsec-eftde}, where we will discuss well-motivated conjectures about which quadrants are allowed or excluded from theoretical predictions, and the relative prior volume of the allowed quadrants.

To calculate cosmological predictions for these phenomenological parametrisations, we use the MGCLASS II code \cite{Sakr:2021ylx} derived from CLASS \cite{Lesgourgues:2011re,Blas:2011rf} and MGCLASS \cite{Baker:2015bva}.

\subsection{$\alpha_i$: Theory-informed parametrisations}\label{subsec-eftde}

As stated in the introduction, in the EFTDE framework the background and linear perturbations are completely described by $w(a)$ (for the background) and the four $\{\alpha_i(a)\}$ (for the perturbations). Out of the $\{\alpha_i\}$, $\alpha_K$ has a negligible effect on subhorizon scales, which is where most of the constraining power of current probes lies. Thus, $\alpha_K$ is very weakly constrained by cosmology \cite{Bellini:2015xja}, and we can safely fix it to a fiducial value when computing our predictions.\footnote{In practice this is often fixed to a positive value smaller than $\mathcal{O}(1)$. The positivity is helpful for avoiding ghost instabilities that arise from $D \equiv \alpha_K + \frac{3}{2}\alpha_B^2 < 0$ \cite{Bellini:2014fua}, and the smallness is helpful for avoiding the sound horizon of the scalar field being much smaller than the Hubble horizon $H^{-1}$, which would cause a breakdown of the quasistatic approximation (explained in \cref{sec-mapping}) \cite{Sawicki:2015zya}. See \cref{subsec-implemdetails} for the precise values used in our analysis. \label{foot_alphaK}}. Also, $\alpha_T$ is frequently fixed to zero in cosmological analyses motivated by measurements of the speed of gravitational waves extremely close to the speed of light following the binary neutron star merger GW170817.
We will discuss the caveats of applying this constraint to cosmological analyses and show the effect of freeing up $\alpha_T$ on the constraints on EFTDE parameters in \cref{sec-GWpriors}. In the rest of the paper, we apply this constraint, and thus focus on ${\alpha_B, \alpha_M}$ which remain as the relevant degrees of freedom for linear perturbations.

In the same spirit as \cref{mu_OmegaDE,gamma_OmegaDE}, the baseline parametrisation that we mainly focus on in this paper (first applied to the $\{\alpha_i\}$ in \cite{Bellini:2014fua,Bellini:2015xja}) is
\begin{gather}
\alpha_B(a) = \alpha_{B,0}\frac{\Omega_{DE}(a)}{\Omega_{DE,0}}\ \nonumber,\\ 
\alpha_M(a) = \alpha_{M,0}\frac{\Omega_{DE}(a)}{\Omega_{DE,0}}\ ,
\label{eqn:baseline_alphas}
\end{gather}
with $\alpha_T$ fixed to zero and $\alpha_K$ fixed to a fiducial value as discussed above. Note that for obtaining the complete evolution of the Planck mass, $M(a)$, one needs to specify $\alpha_M(a) = d \ln M /d \ln a$ as well as the value of $M^2$ at any single instant. We fix $M \rightarrow 1$ as $a \rightarrow 0$ with the purpose of making deviations from GR vanish at early times. We report constraints on the present day values $\alpha_{B,0}, \alpha_{M,0}$. This makes it straightforward to compare constraints on these quantities across different parametrisations for the $\alpha_i$ and associated different time dependences, as we do in \cref{sec-scale,sec-shiftsym}.

\togglered{Similarly to the phenomenological $\{\mu, \gamma\}$ parametrisation, while we assume the time dependence of \cref{eqn:baseline_alphas} as the fiducial for most of our analyses, another phenomenological time dependence considered for $\{\alpha_B,\alpha_M\}$ is}
\begin{gather} \label{alphaB_scale}
\togglered{\alpha_B(a) = \alpha_{B,0}a\ ,}\\ \label{alphaM_scale}
\togglered{\alpha_M(a) = \alpha_{M,0}a\ ,}
\end{gather}
\togglered{which we compare with \cref{eqn:baseline_alphas} in detail in \cref{sec-scale}.}

For any parametrisation of the time dependence, a substantial restriction on the present day values $\alpha_{B,0}, \alpha_{M,0}$ comes from the requirement that the squared sound speed of the scalar field must be positive for the controlled growth of perturbations. This is known as the gradient stability requirement. The expression for the sound speed is
\begin{gather} \label{cs2}
c_s^2(a) = \dfrac{1}{D}\left\{(2 - \alpha_B)\left(\hat{\alpha} - \dfrac{\dot{H}}{H^2}\right) - \dfrac{3(\rho_m + p_m)}{H^2M^2} + \dfrac{\dot{\alpha_B}}{H}\right\}\ ,
\end{gather}
where we have suppressed scale factor arguments on the RHS for clarity, $D \equiv \alpha_K + \frac{3}{2}\alpha_B^2$, and $\hat{\alpha} \equiv \frac{1}{2}\alpha_B(1+\alpha_T) + \alpha_M - \alpha_T$. 
Since we have fixed $\alpha_K$ to be positive (see \cref{foot_alphaK}) and $D$ is always positive to avoid the ghost instability, the sign of $c_s^2$ is determined by an interplay between the $\{\alpha_i(a), w(a)\}$ except $\alpha_K(a)$. Note that the gradient stability requirement can in principle be violated for short periods at early times (potentially as an artefact of the parametrisations chosen) without significantly affecting observables -- this does not substantially alter the allowed parameter space of $\BMz$ for most of the parametrisations that we consider \togglered{(as checked by \cite{Noller:2018wyv})} and we therefore simply present bounds where the gradient stability criterion is strictly enforced. The single exception is the gravitational wave prior motivated case \cref{GWprior_omegaDE}, which we discuss in \cref{sec-GWpriors}. \togglered{For this model, mild gradient instabilities at early times are an artefact of the $\propto \Omega_{DE}$ time dependence, and imposing a strict gradient stability requirement falsely excludes a significant amount of viable parameter space. This can be dealt with by adding a small term to the kineticity in order to satisfy gradient stability without affecting any observables, an approach discussed in more detail in \cite{Noller:2020afd}.}

While \cref{eqn:baseline_alphas} represents a theoretically motivated set of basis functions (controlling the dynamics of linear perturbations), their time dependence is phenomenologically parametrised. In \cref{sec-shiftsym} we will discuss in more detail what effect adding a theoretically motivated time dependence has on the resulting constraints.

We use the hi\_class code \cite{Zumalacarregui:2016pph,Bellini:2019syt}
to calculate cosmological predictions for these models.

\section{Mapping between parametrisations}\label{sec-mapping}

In EFTDE models, the relation between the gravitational potential and the density perturbations reduces to a form captured by \cref{Poisson_mu,Poisson_Sigma} (with specific expressions for $\mu$ and $\Sigma$ in terms of the underlying EFT parameters, which we will see below) when: 1) the time derivative of the scalar field can be neglected compared to its spatial derivatives (this approximation is known as the quasi-static approximation or QSA), and 2) the scale-dependence in the growth of matter perturbations can be neglected. For the QSA, a necessary condition for it to be valid is that the scales being probed are much smaller than the sound horizon of the scalar field. While for scale-independent growth, the scale associated with the mass of the scalar field, $m$, needs to be comparable to (or larger than) the Hubble horizon $H^{-1}$. The sound horizon of the scalar field is comparable to the Hubble horizon unless $c_s^2 \ll 1$, which in the models we consider only occurs for a small fraction of the parameter space that is close to violating the gradient stability condition. \togglered{We confirm this expectation in \cref{app-cs2} where we show the values of $c_s^2$ at $z=0$ for two models with different time-dependences of $\{\alpha_B, \alpha_M\}$.} The Compton wavelength, $m^{-1}$, is also comparable to $H^{-1}$ in our models because self-accelerating models (i.e. models where the cosmological constant is absent and accelerated expansion is caused solely by the dynamics of the scalar field) have the mass of the field comparable to the horizon scale (or larger), and here we only consider self-accelerating EFTDE models. 

Thus, both the QSA and scale-independent growth are excellent approximations on subhorizon scales for the models we are considering here. With most of the constraining power of our probes coming from sub-horizon scales, we can therefore employ the QSA to accurately recover the full evolution of gravitational perturbations \cite{Sawicki:2015zya}, where we point out that the accuracy of the QSA can be related to the proximity of the background evolution to that of $\Lambda{}$CDM \cite{delaCruz-Dombriz:2008ium,Noller:2013wca}. Usefully, the QSA allows us to write a simple analytic form of the Poisson equations for the two gravitational potentials, and scale-independent growth allows us to neglect the scale dependence of the modifications to gravity, leading to \cref{Poisson_mu,Poisson_Sigma} also holding in EFTDE models. Under these two approximations, and further assuming $\alpha_T = 0$ (an assumption that we will relax in \cref{sec-GWpriors}), the mapping from the EFTDE functions to $\mu(a), \Sigma(a)$ on subhorizon scales is given by

\begin{gather} \label{mu_qsa}
\mu(a) = \dfrac{1}{M^2}\left(1 + \dfrac{(\alpha_B + 2\alpha_M)^2}{2Dc_s^2}\right)\ ,\\ \label{Sigma_qsa}
\Sigma(a) = \dfrac{1}{M^2}\left(1 + \dfrac{(\alpha_B + \alpha_M)(\alpha_B + 2\alpha_M)}{2Dc_s^2}\right)\ .
\end{gather}

\begin{figure}[t]
    \centering
    \includegraphics[width=\linewidth]{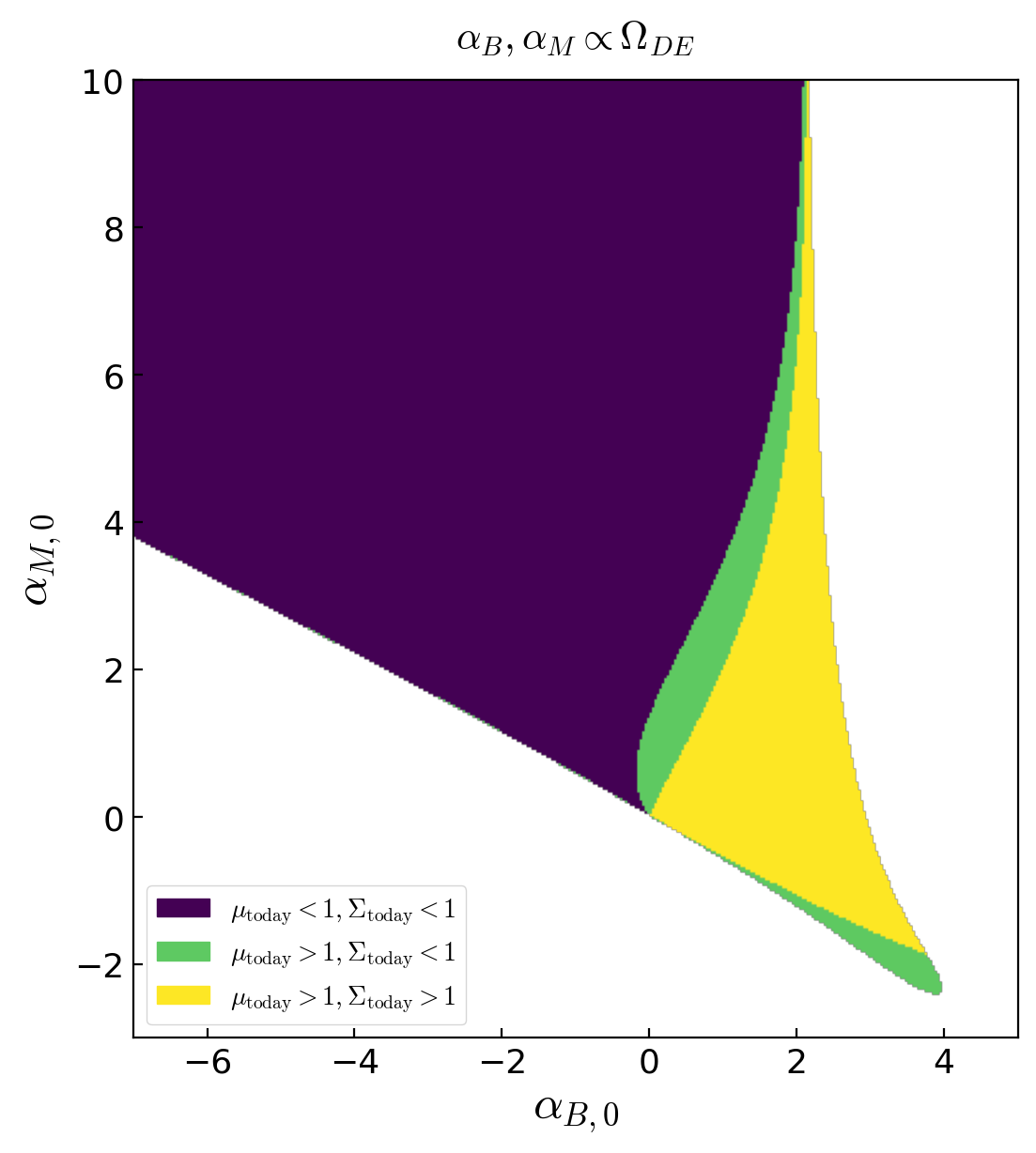}
    \caption{The allowed parameter space for $\BMz$ for the EFTDE parametrisation $\alpha_B, \alpha_M \propto \Omega_{DE}$ (explicitly, \cref{eqn:baseline_alphas}). The coloured (white) regions correspond to $\BMz$ values where the gradient stability condition $c_s^2 > 0$ is obeyed (violated) today. The bounds at $\alpha_{M,0}=10$ and $\alpha_{B,0}=-7$ are imposed by hand to limit the size of the region, while the lower bound on $\alpha_{M,0}$ and upper bound on $\alpha_{B,0}$ are physical boundaries imposed by the gradient stability condition. The colours indicate the quadrant of $\mSt$ space that a particular $\BMz$ maps to from evaluating \cref{mu_qsa,Sigma_qsa} at $a=1$. Note the different prior volumes of the quadrants: in particular, the $\mt > 1, \St < 1$ quadrant has the smallest prior volume, in accordance with the conjecture of \cite{Pogosian:2016ji}. The image of this mapping in the $\mSt$ space is shown in \cref{mapping_omegaDE}.}
    \label{musigma_signs_omegaDE}
\end{figure}

\begin{figure}[t]
    \centering
    \includegraphics[width=\linewidth]{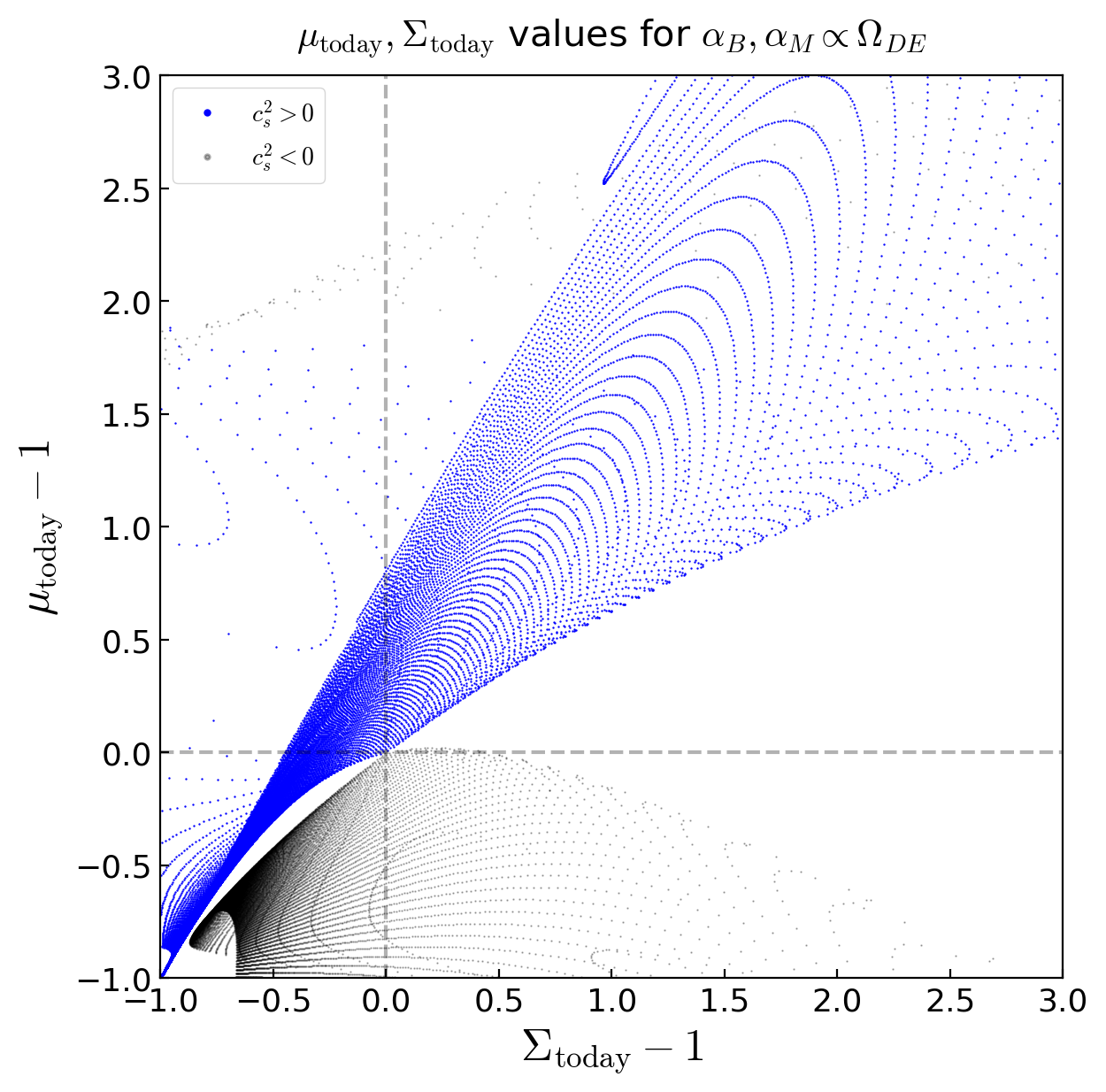}
    \caption{The image of the theoretical parameter space in \cref{musigma_signs_omegaDE} in the $\mSt$ space under the mapping \cref{mu_qsa,Sigma_qsa}. The points shown here are images of a uniform sampling of the $\BMz$ parameter space enclosed in \cref{musigma_signs_omegaDE}. The axis ranges are limited for clarity. Blue (grey) points are derived from $\alpha_{B,0}, \alpha_{M,0}$ values where the gradient stability condition $c_s^2 > 0$ is obeyed (violated) today, which corresponds to the coloured (white) regions in \cref{musigma_signs_omegaDE}. Note the clear separation between the stable and unstable $\mSt$ parameter spaces, and the fact that even the combination of stable and unstable parameter spaces does not cover the entire $\mSt$ plane. Both these features are due to the mathematical nature of the nonlinear mapping from the $\BMz$ space to the $\mSt$ space, \cref{mu_qsa,Sigma_qsa}.}
    \label{mapping_omegaDE}
\end{figure}

We use the present day values $\mSt$ defined in \cref{mu_today,Sigma_today} as proxies for their time-averaged effect in the late universe. It is straightforward to derive these by evaluating \cref{mu_qsa,Sigma_qsa} at $a=1$. We show the allowed parameter space for ${\BMz}$ (limited only by the gradient stability condition $c_s^2 > 0$), and its correspondence with the four quadrants $\mt < 1 / > 1, \St < 1 / > 1$ in \cref{musigma_signs_omegaDE}. In \cref{mapping_omegaDE}, we show how this parameter space maps to the space of $\mSt$ values.

When we choose to be completely agnostic about the values of $\BMz$ (meaning we have wide uniform priors on both of them), \cref{musigma_signs_omegaDE} shows the relative prior volumes of the four quadrants $\mt < 1 / > 1, \St < 1 / > 1$ in the $\mSt$ space. In \cite{Pogosian:2016ji}, it was conjectured that in any theories within the Horndeski class of the EFTDE, the two quadrants where $\{\mt - 1, \St - 1\}$ have opposite signs should have a much smaller prior volume than the other two quadrants where they have the same sign. \cref{musigma_signs_omegaDE} shows a clear example for which this conjecture holds: in fact, the $\mt < 1, \St > 1$ quadrant is entirely absent from the predictions of the stable theory. The conditions for a theory to allow this quadrant have been studied in \cite{Peirone:2017ywi}.

The gradient stability requirement imposes a hard upper bound on $\alpha_{B,0}$ at $\alpha_{B,0}\  \lsim\  4$ and a hard lower bound on $\alpha_{M,0}$ at $\alpha_{M,0}\  \gsim -2.5$. This is useful when setting priors on $\BMz$ in cosmological analyses of the $\alpha_i \propto \Omega_{DE}$ model, as these bounds serve as physically motivated prior boundaries. However, these bounds are subject to change if the background expansion history is varied, since it also enters the expression for $c_s^2$.

There is no physical upper bound on $\alpha_{M,0}$ or lower bound on $\alpha_{B,0}$, so in principle the total allowed prior volume in the $\BMz$ space is infinite. However, when considering the $\mSt$ space, \cref{musigma_signs_omegaDE} clearly shows that extending the two non-physical boundaries in the $\BMz$ space beyond a certain point has a very limited effect. By extending the upper bound on $\alpha_{M,0}$ over that in \cref{musigma_signs_omegaDE}, or the lower bound on $\alpha_{B,0}$ (which remains physically viable only if the upper bound on $\alpha_{M,0}$ is sufficiently high), we only add prior volume to the $\mt < 1, \St < 1$ quadrant. This is because as $\alpha_{M,0}$ becomes larger, the Planck mass today increases exponentially. This exponentially suppresses all gravitational interactions of cosmological perturbations, which corresponds to $\mt, \St \rightarrow 0$. Thus, for a prior on $\alpha_{M,0}$ unbounded from above, both $\mt > 1, \St < 1$ and $\mt > 1, \St > 1$ have a negligible fraction of the volume in the $\alpha_{B,0}, \alpha_{M,0}$ parameter space. A completely uninformative prior on $\alpha_{M,0}$ translates to an extremely informative prior on $\mt, \St$ preferring $\mt, \St \rightarrow 0$, which is very unrealistic from an observational standpoint as it would lead to no growth of structure.

In \cref{mapping_omegaDE}, we show the range of $\mSt$ values that can be attained from the EFTDE parameter space for the model \cref{eqn:baseline_alphas} shown in \cref{musigma_signs_omegaDE}. First, we focus on the stable parameter space of the theory, which is represented by the blue points in \cref{mapping_omegaDE}. It can be clearly seen that the requirement of deriving $\mSt$ from an underlying EFTDE imposes a theoretical prior in the $\mSt$ space with both the prior density and prior boundaries being nontrivial.

The stable region in \cref{mapping_omegaDE} has clear boundaries. It should be stressed that they do not correspond to the saturation of the gradient stability condition ($c_s^2 = 0$), because this saturation corresponds to infinite values of $\mSt$ as can be seen from \cref{mu_qsa,Sigma_qsa}. Rather, these boundaries arise from the nature of the nonlinear mapping of \cref{mu_qsa,Sigma_qsa} (we also note that the upper left boundary of this region is permeable in principle, but with a sharp decrease in the density of the mapping).

Our prior in the $\alpha_{B,0}, \alpha_{M,0}$ space is uniform. However, the variation of the density of points in \cref{mapping_omegaDE} shows that the prior imposed from the EFTDE on $\mSt$ within the allowed ranges of their values is not uniform. Thus, the volume enclosed by the boundaries in any quadrant, or generally any region of \cref{mapping_omegaDE} does not accurately reflect the total prior volume in that region (which directly follows from the fact that \cref{mu_qsa,Sigma_qsa} are not linear in $\BMz$.). The correct representation of the prior volume of a region is the total number of points enclosed in the region, which is because our prior on $\alpha_{B,0}, \alpha_{M,0}$ is uniform, so that each point in the $\mSt$ space has the same weight as it corresponds to a single point in the $\BMz$ space. The prior volumes of the various quadrants in \cref{mapping_omegaDE} are more easily seen from \cref{musigma_signs_omegaDE}.

There are no points with $c_s^2 > 0$ that have $\mt < 1, \St > 1$. The mathematical conditions to have opposing signs of ${\mu - 1, \Sigma - 1}$ were discussed in \cite{Peirone:2017ywi}, where it was found that in order to have $\mt < 1, \St > 1$ and satisfy gradient stability, a necessary condition is $\alpha_M(\alpha_B + 2\alpha_M) < 0$\footnote{This is Eq. 36 in \cite{Peirone:2017ywi}, where we have made the substitution $g_3 = - (\alpha_B + \alpha_M)$.} (this simply follows from imposing $\mu < \Sigma$ and $Dc_s^2 > 0$ in \cref{mu_qsa,Sigma_qsa}). For parametrisations where $\alpha_M$ varies monotonically and GR is recovered at early times, we cannot have $\mu < 1$ if $\alpha_M < 0$ from \cref{mu_qsa} (as we have assumed $M=1$ at early times and $M$ needs to increase to have $\mu < 1$). Thus, the necessary condition becomes
\begin{gather}\label{musigma_fourthquadrant}
\alpha_M > 0,\ \alpha_B + 2\alpha_M < 0\ .
\end{gather}
This is a very substantial restriction on the parameter space, and in the case of the $\alpha_i \propto \Omega_{DE}$ parametrisation, it is ruled out by the gradient stability condition (at least at $a=1$, from \cref{musigma_signs_omegaDE}). We have thus shown that for this parametrisation, $\alpha_{M,0} > 0$ rules out not only $\mt < 1, \St > 1$, but also the weaker condition $\mt < \St$.

Let us now focus on the unstable parameter space of the theory, represented by the grey points in \cref{mapping_omegaDE} which are mapped from the white region in \cref{musigma_signs_omegaDE}. Note that when there is a gradient instability, the QSA would also break down, so we cannot read much into the precise values of $\mSt$ given by the QSA expressions \cref{mu_qsa,Sigma_qsa} for unstable regions of $\alpha_{B,0}, \alpha_{M,0}$. Despite this fact, plotting these QSA values is informative. In \cref{mapping_omegaDE}, the stable and unstable parameter spaces for $\mSt$ are clearly separated, with a wide gap between them that is inaccessible via the mapping \cref{mu_qsa,Sigma_qsa}. We can also conclude that the boundaries of the stable and unstable parameter spaces do not arise from the saturation of the gradient stability condition, $c_s^2 = 0$, because $c_s^2$ appears in the denominators in \cref{mu_qsa,Sigma_qsa}. Thus, points in the $\BMz$ space that saturate the gradient stability condition are not mapped to the $\mSt$ space, since their $\mSt$ values are undefined. Rather, the boundaries of the stable and unstable $\mSt$ parameter spaces in \cref{mapping_omegaDE} arise from the mathematical nature of the highly nonlinear mapping of \cref{mu_qsa,Sigma_qsa} from the $\BMz$ space to the $\mSt$ space.

\section{Data sets and likelihood implementation}\label{sec-data}

In this section, we describe the various probes that we use to constrain modified gravity and other cosmological parameters. We also list the analysis choices implemented in the likelihoods.

\subsection{Datasets}

\begin{itemize}

\item {CMB}: We use the official pre-marginalised Planck 2018 likelihoods, which are composed of \emph{plik-lite} (TTTEEE, $\ell \geq 30$), \emph{Commander} (TT, $\ell < 30$), \emph{SimAll} (EE, $\ell < 30$), and the CMB lensing likelihood \cite{Planck:2018lbu,Planck:2018vyg,Planck:2019nip}. The two main gravitational effects on the CMB in the late universe are gravitational lensing and the late Integrated Sachs-Wolfe (ISW) efect. These effects are detectable in the CMB lensing and low-$\ell$ TT likelihoods respectively, which is why these two likelihoods contribute the most constraining power for late-universe modified gravity. The other CMB likelihoods are important for tightly constraining the remaining cosmological parameters independently of the late-time modified gravity effects on perturbations.

The CMB lensing likelihood is roughly sensitive to redshifts $0\ \lsim \ z\ \lsim \ 2$, based on Figure 4 of \cite{Planck:2018vyg} which shows the redshift sensitivities of different multipoles of the lensing potential power spectrum $C_{\ell}^{\phi\phi}$. It should be noted that there is no sharp cutoff of redshift sensitivity at $z=2$, but the sensitivities fall off gradually beyond this redshift.

The CMB likelihoods come with the nuisance parameter $A_L$, which is defined as an artificial rescaling of the lensing potential power spectrum via 
\begin{gather}
C_\ell^{\Psi_L\Psi_L} \rightarrow A_LC_\ell^{\Psi_L\Psi_L}\ ,
\end{gather}
where the lensing potential $\Psi_L$ is given by the combination of gravitational potentials $\Phi + \Psi$ appearing in \cref{Poisson_Sigma} projected along the line of sight with a lensing kernel. A preference for $A_L \neq 1$ is by definition unphysical, and points to either new physics or to unaccounted for systematics in the likelihoods. Since the newer Planck PR4 likelihoods of \cite{Tristram:2023haj} find $A_L$ consistent with unity within $\Lambda$CDM, it indicates that there may be remaining systematics in the older official Planck 2018 likelihoods that we use. Marginalising over $A_L$ is then a conservative choice to account for these systematics and this is the choice that we adopt here. Another motivation for being conservative about $A_L$ is that it is degenerate with modified gravity effects on lensing, so a preference for $A_L \neq 1$ can correspond to a preference for modified gravity when $A_L$ is fixed to unity. This degeneracy is well-studied \cite{Pogosian:2021mcs, Specogna:2024euz}, however the strength of the degeneracy and its impact on modified gravity constraints depends on the model. In \cref{app-AL}, we show the impact of this choice on modified gravity constraints on various models by comparing constraints obtained with varying $A_L$ with those obtained from fixing $A_L=1$.

\item {Galaxy Clustering}: Galaxy clustering observations measure the growth rate of structure using redshift space distortions (RSD), and the expansion rate of the universe using baryon acoustic oscillations (BAO). In reality, the RSD and BAO measurements are correlated. We use the joint RSD+BAO likelihoods from the eBOSS DR16 compilation \cite{eBOSS:2020yzd, BOSS:2016wmc, eBOSS:2020hur, eBOSS:2020uxp, eBOSS:2020tmo} which also includes measurements from BOSS DR12 \cite{BOSS:2016wmc} and MGS \cite{Howlett:2014opa}, consistently combining RSD and BAO taking their covariance into account. These are in the form of joint measurements of $D_H$, $D_V$, and $f\sigma_8$ at various effective redshifts. The redshift sensitivity of the RSD observations is approximately $0.15\ \lsim \ z\ \lsim \ 1.48$, which is the range of the effective redshifts of the bins in eBOSS DR16 that have RSD measurements \cite{eBOSS:2020yzd}. The BAO observations are sensitive to higher redshifts, up to $z \simeq 2.33$.

A potential concern for using these measurements to constrain modified gravity models is that they employ a $\Lambda$CDM template for the $f\sigma_8$ reconstruction. However, \cite{Barreira:2016ovx} have shown that such growth rate estimates are unbiased for modified gravity models that have a scale-independent modification to the growth rate (at least for current observations), which holds for the EFTDE models we consider as explained in \cref{sec-mapping}.

\togglered{While our focus is on constraining modified gravity perturbations in most of this paper, in \cref{sec-background} we focus on the background expansion. There, in addition to showing background constraints with the eBOSS data, we also show constraints where we replace eBOSS data with the recent observations from DESI DR2 BAO \cite{DESI:2025zgx} in order to study whether the hints for deviation from a $\Lambda{}$CDM background seen by DESI remain for an EFTDE model which imposes a theoretical prior on $\wzwa$.}

\item{SNe}: As Type IA supernovae are standardisable candles, measuring their apparent luminosity across a range redshifts lets us infer the expansion history in that range of redshifts. We use the Pantheon+ likelihood \cite{Pan-STARRS1:2017jku}, which tightly constrains the cosmic expansion history using supernovae from redshifts $0.01\ \lsim \ z\ \lsim \ 2.26$. It comes with a nuisance parameter $M$, which is the absolute magnitude of the supernovae. It does not affect expansion history constraints (as they only impact apparent magnitudes), but it is degenerate with $H_0$ when using supernova data alone. In our case, we vary $M$ with a wide uniform prior, and its posterior is determined from the tight constraints on $H_0$ from our CMB and BAO datasets.

\item {ISW}: The ISW effect on the CMB temperature autocorrelations is an integrated effect over all redshifts from $z=0$ to the redshift at last scattering. The detection of its significance, and thereby its constraining power, is highly limited by cosmic variance. By using measurements of the cross-correlations of the CMB temperature anisotropy with galaxy number counts that have photometric redshift measurements, we can separate the contributions of the ISW effect into different redshift bins and gain constraining power. The biases of the various galaxy samples are nuisance parameters for the ISW likelihood. The likelihood thus also uses the autocorrelations of the galaxy number counts, which break the degeneracy between galaxy bias and the ISW contribution. We use the ISW likelihood from \cite{Seraille:2024beb}, which is an adaptation of the ISW likelihood of \cite{Stolzner:2017ged} to modified gravity models. The likelihood uses number counts and photometric redshifts from the 2MPZ \cite{Bilicki:2013sza}, WISE $\times$ SuperCOSMOS \cite{Bilicki:2016irk}, SDSS-DR12 photo-z \cite{Beck_2016}, SDSS-DR6 QSO \cite{Richards:2008eq}, and NVSS \cite{Condon:1998iy} catalogs. We also adopt a simplification of the bias modelling introduced and justified in \cite{Seraille:2024beb}: namely we fix the redshift evolution of the bias across all the bins of a particular survey, thus reducing the set of nuisance parameters to one bias amplitude per survey instead of one bias amplitude per survey bin.

The approximate redshift sensitivity of the ISW likelihood is $0\ \lsim \ z\ \lsim \ 2$. This is inferred from the ranges of the peaks of the photometric redshift distributions of the various surveys shown in Figure 1 of \cite{Stolzner:2017ged}. Like for the CMB lensing likelihood, there isn't a sharp cutoff in sensitivity at $z = 2$, but most of the sources lie within this redshift.

The ISW likelihood computes the CMB-galaxy cross-correlations from the Planck 2015 CMB temperature maps. Using this alongside our Planck 2018 likelihoods is not a source of concern because in \cite{Stolzner:2017ged}, the authors confirmed that the ISW signal varies negligibly across different CMB maps from Planck 2015. Its variation should thus also be negligible when the Planck 2015 maps are updated to Planck 2018.

\end{itemize}

\subsection{Likelihood implementation details}\label{subsec-implemdetails}

We use the MontePython code \cite{Brinckmann:2018cvx,Audren:2012wb} to perform a Markov Chain Monte Carlo inference of the cosmological and nuisance parameters. We vary six standard cosmological parameters: $\{\omega_b, \omega_c, \theta_s, \ln(A_s), n_s, \tau\}$ with uniform priors. For the reionisation optical depth $\tau$, we impose a nontrivial upper bound of $\tau \geq 0.04$ from observations of the Gunn-Peterson trough (see e.g. \cite{SDSS:2001tew}). For our baseline constraints, we vary $A_L$ with a uniform prior $0.7 \leq A_L \leq 1.3$, which is sufficiently wider than the posteriors we obtain for $A_L$ in all our models.

We now list some analysis choices we made for EFTDE models. While the choice of $\alpha_K(z)$ does not affect phenomenology in the regime we are interested in, as discussed in \cref{subsec-eftde}, in practice we have to decide on a particular $\alpha_K(z)$ for running the MCMC analyses. We adopt a range of choices based on previous use in the literature \cite{Noller:2018wyv}: $\alpha_K = 0.1\Omega_{DE}$ for the $\alpha_B, \alpha_M \propto \Omega_{DE}$ model, and $\alpha_K = 0.1a$ for the $\alpha_B, \alpha_M \propto a$ model. For the shift symmetric model of \cref{sec-shiftsym}, we use $\alpha_K = 0.1$ as a constant in redshift, as there is no natural choice for its time dependence.

We make two further analysis choices that mitigate numerical instabilities in hi\_class without affecting phenomenology in the observable regime. First, we add a small constant, $0.01$, to $\alpha_K(z)$ in parametrisations where it decays with redshift: this is to prevent rapidly oscillating modes of the scalar field at very early times that lead to numerical instabilities but are unobservable. Second, we implement a miniscule relaxation of the gradient stability condition: $c_s^2 \geq -10^{-14}$ instead of $c_s^2 \geq 0$. This is to prevent spurious violations of the condition due to floating point precision errors.

For phenomenological modified gravity models, the only analysis choice made is that we impose the prior bound $\gamma > 0$, or equivalently $\Sigma(z) > \mu(z)/2$ to avoid numerical instabilities in MGCLASS. The same prior bound has been used by the DES collaboration \cite{DES:2022ccp} which used MGCAMB. All our posteriors are far from this boundary.

We define the convergence criterion of our MCMC chains as $R - 1 < 0.01$, where $R$ is the Gelman-Rubin convergence criterion computed from the chains after discarding the burn-in. This criterion is satisfied by the chains of most of our models, with the only exceptions being the two shift-symmetric models in \cref{sec-shiftsym} (one with a $\Lambda$CDM background and one with a CPL background)\togglered{, and the EFTDE run with DESI DR2 BAO shown in \cref{w0wa_comparison_desi}}, for which we stopped at $R - 1 < 0.02$ due to not seeing further improvement with longer chains.

Finally, our emphasis in this paper is on the qualitative impact of various theoretical priors on the observational posteriors of dynamical DE functions, and not on the precise magnitudes of the constraints on the specific parameters. For this reason, we only show the graphical posteriors on parameters in the following sections, and refer the reader to \cref{tab:parameter_constraints} where we have collected the precise constraints on parameters derived from these posteriors. We use the GetDist code \cite{Lewis:2019xzd} to plot observational posteriors and derive the parameter constraints in \cref{tab:parameter_constraints}.

\begin{figure*}[t]
    \centering
    \begin{subfigure}
        \centering
        \includegraphics[width=0.51\linewidth]{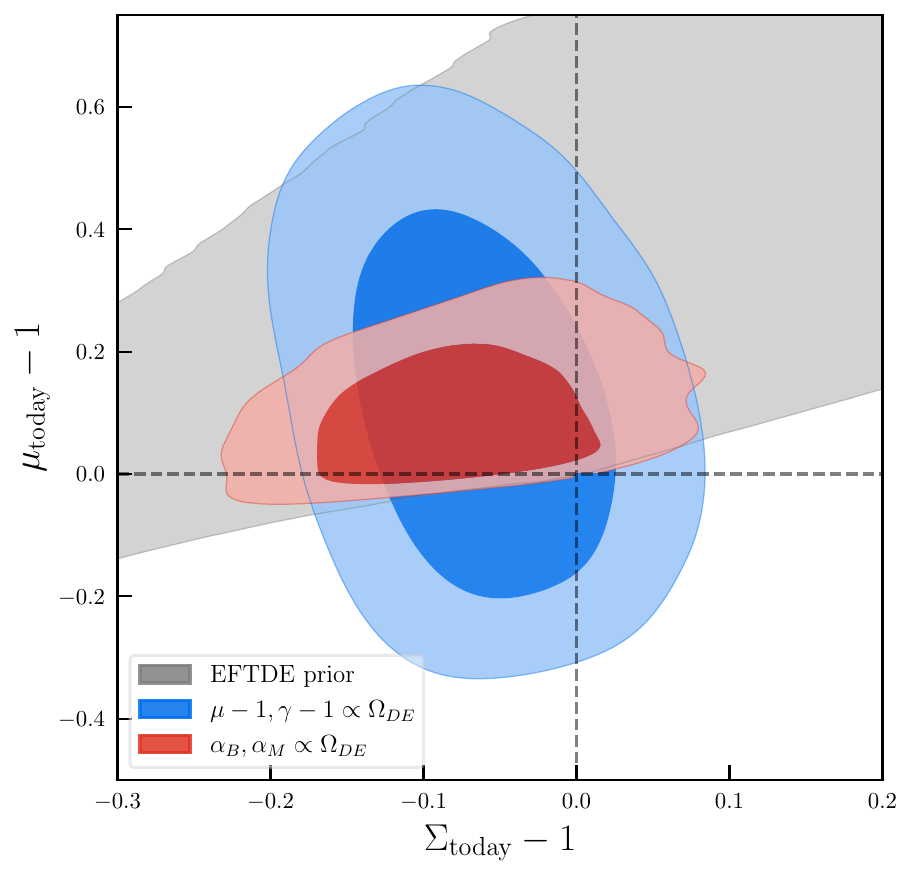}
    \end{subfigure}
    \begin{subfigure}
        \centering
        \includegraphics[width=0.48\linewidth]{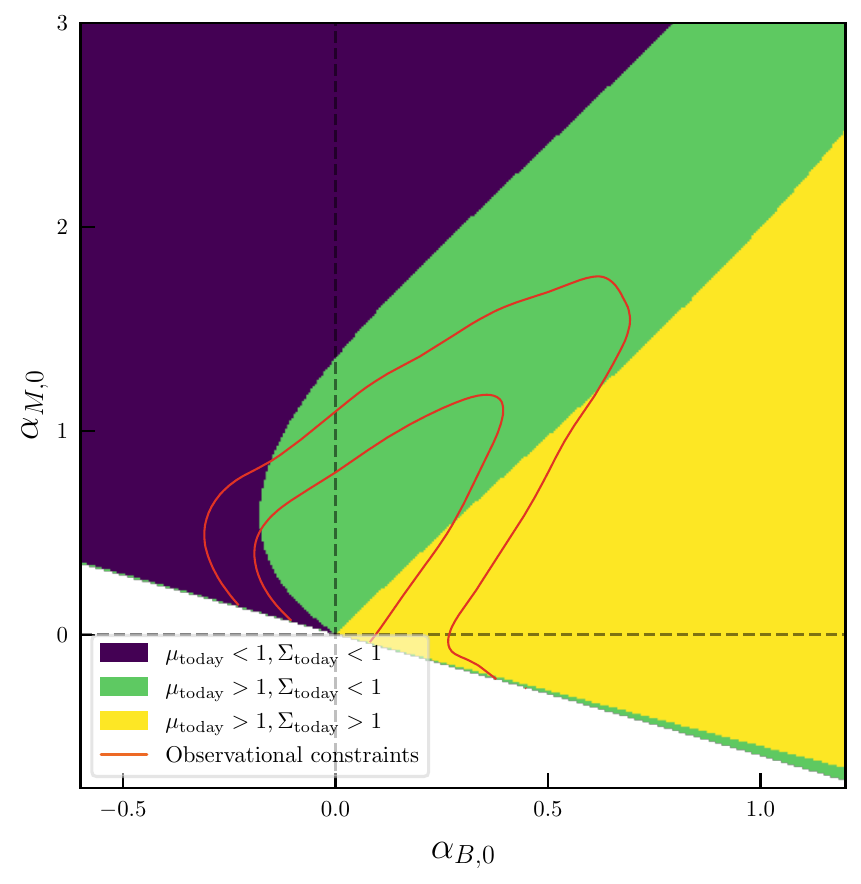}
    \end{subfigure}
    \caption{\textbf{Left panel:} Comparison of observational constraints on $\mSt$ from phenomenological and EFTDE parametrisations. The grey contour shows the allowed parameter space of the EFTDE prior, with its boundaries arising from the mapping \cref{mu_qsa,Sigma_qsa}. We only show the prior boundaries and not the density of points in the prior, as the relative density of different regions in the prior depends on the prior on the EFTDE parameters as explained in \cref{sec-mapping}. Working with a theory-informed EFTDE parameter basis clearly results in significantly tighter bounds, which is the combined consequence of the EFTDE prior shown here, and the different time evolution of $\{\mu, \Sigma\}$ shown in \cref{musigma_evolution_omegaDE}. \textbf{Right panel:} Comparison of observational constraints with the theoretically allowed parameter space for the $\alpha_B, \alpha_M \propto \Omega_{DE}$ model. An interesting feature of this posterior is that the $\mt > 1, \St < 1$ region which has the smallest relative volume in the prior space has the largest relative volume in the posterior.}
    \label{omegaDE_theory_observations}
\end{figure*}

\section{Theory prior I: Phenomenological vs. EFT parametrisations}\label{sec-phenoEFT}

In \cref{sec-mapping}, we found that the mapping from the EFTDE parameters $\BMz$ to the phenomenological parameters $\mSt$ restricts the allowed values of $\mSt$ compared to the phenomenological parametrisation, where all values are allowed. This is a restriction in the prior volume, i.e. without any constraints from data. In this section, we show the difference in the posteriors of $\mSt$ between the two parametrisations. The difference in the posteriors arises from both the difference in priors and the constraints on $\BMz$ from data.

In \cref{omegaDE_theory_observations}, we show the theoretical priors and observational posteriors in both the $\mSt$ and $\BMz$ spaces. Considering first the phenomenological parametrisation \cref{mu_OmegaDE,gamma_OmegaDE}, there is no theoretical prior on $\mSt$. We find that the observational 2$\sigma$ C. L. contours span all the four quadrants around $\mSt = \{1,1\}$. The GR limit lies within the 1$\sigma$ contours.

Now we consider the EFTDE parametrisation \cref{eqn:baseline_alphas}. Looking at the $\mSt$ space, \cref{omegaDE_theory_observations} shows the bounds of the theoretical prior on $\mSt$ shown in \cref{mapping_omegaDE}. These bounds are drawn by sampling a large prior volume\footnote{We use the emcee code \cite{Foreman-Mackey:2012any} for this sampling. We fix the other cosmological parameters to their Planck 2018 best fit values for this sampling. Their values have a minimal effect on the allowed parameter space of $\BMz$.} in the $\BMz$ space which is shown in the right panel of \cref{omegaDE_theory_observations}, and mapping it to the $\mSt$ space via \cref{mu_qsa,Sigma_qsa}. The roughness at the upper boundary of the prior results from the sparsity of the mapping in that region. The theoretical prior cuts off a significant portion of the allowed posterior with $\mt < 1$. It completely forbids $\mt < 1, \St > 1$, for reasons discussed in \cref{sec-mapping}, and highly restricts the $\mt < 1, \St < 1$ region by not allowing the ratio $\mt/\St$ to be very small (intuitively, the mathematical conditions required for a low value of $\mu/\Sigma$ should be closely related to the conditions required for $\mu < \Sigma$, which are discussed in \cref{sec-mapping}). Note that the marginal crossing of the theoretical prior by the EFTDE posterior in the $\mt > 1, \St > 1$ quadrant is not a smoothing artefact, but it is due to the fact that the theoretical prior bounds fluctuate very slightly depending on the value of $\Omega_{m,0}$, which was held fixed for drawing the bounds in the figure.

Looking at the $\BMz$ space, we find something interesting: the observational posterior prefers the values of $\BMz$ for which $\mt > 1, \St < 1$, which is the quadrant with the smallest prior volume. This is a substantial caveat to the conjecture of \cite{Pogosian:2016ji, Peirone:2017ywi} about the prior volumes of the various quadrants: while the $\mt > 1, \St < 1$ quadrant has the least volume in the prior, this effect is not enough to overcome the constraining power of the likelihoods. Thus, this quadrant has the highest volume in the posterior despite having the smallest volume in the prior. It is encouraging to find that even with datasets prior to Stage IV LSS surveys, our likelihoods are constraining enough to overcome prior volume effects in the $\BMz$ space. The combination of probes of clustering and lensing is crucial to achieve this level of constraining power.

\begin{figure*}[t]
    \centering
    \begin{subfigure}
        \centering
        \includegraphics[width=0.48\linewidth]{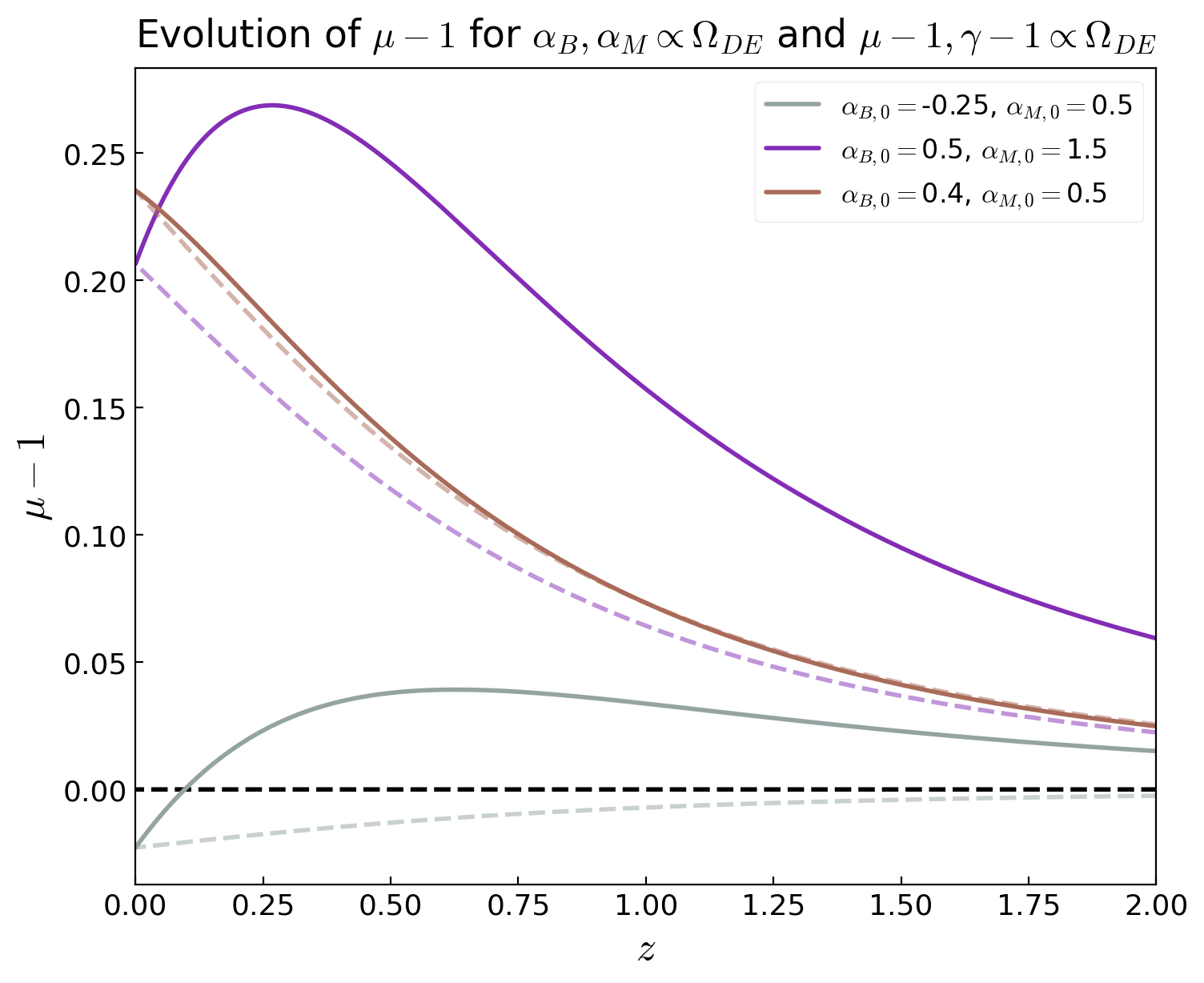}
    \end{subfigure}
    \begin{subfigure}
        \centering
        \includegraphics[width=0.49\linewidth]{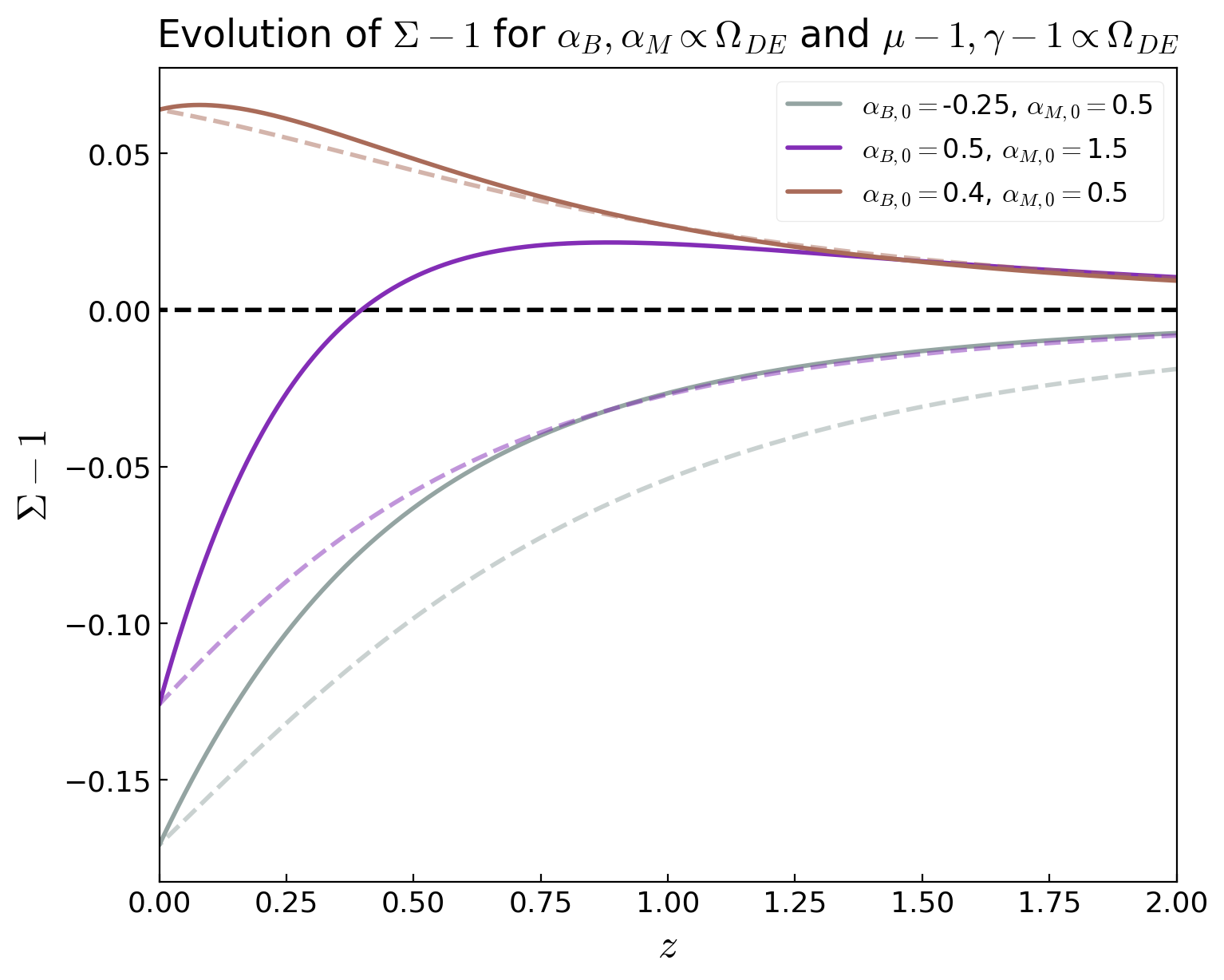}
        
    \end{subfigure}
    \caption{The time evolution of $\mu$ and $\Sigma$ for EFTDE and phenomenological parametrisations. The solid lines are the time evolution in the EFTDE parametrisations, while the faint dashed lines are the time evolution in the phenomenological parametrisations with the same $\mt, \St$ values. We show the time evolutions for one point in each of the three observationally allowed quadrants in the $\mSt$ space for EFTDE parametrisations (see \cref{sec-mapping}). The phenomenological parametrisation is not able to reproduce features seen in the EFTDE parametrisation such as an increase in $\mu$ at higher redshifts, or a sign change of $\mu - 1$ or $\Sigma - 1$.}
    \label{musigma_evolution_omegaDE}
\end{figure*}

Looking again in the $\mSt$ space, the posterior for the EFTDE parametrisation is much narrower than the one for the phenomenological parametrisation. The theoretical prior is too wide to explain the entire difference in posteriors, especially for $\mt > 1$. To fully understand this difference, we need to understand how structure growth and lensing differ between the two parametrisations for the same values of $\mSt$ and the same primordial perturbations. This is directly related to how the past evolutions of $\{\mu, \Sigma\}$ differ between the two parametrisations for the same $\mSt$ values, which is shown in \cref{musigma_evolution_omegaDE} for some particular models which lie within the observational posteriors of \cref{omegaDE_theory_observations}.

From \cref{musigma_evolution_omegaDE} we see how accurately the phenomenological parametrisation can mimic the modifications to gravity from an underlying EFTDE in various regimes. We find that for models with $\mt, \St > 1$, the phenomenological and EFTDE models have a similar time dependence of $\mu, \Sigma$. For models with $\mt > 1, \St < 1$, the EFTDE model has stronger enhancement of growth, while the suppression of lensing is weaker at low redshifts, and changes to enhancement of lensing at high redshifts. And for models with $\mt, \St < 1$, the EFTDE models have weaker suppression of lensing, and enhanced rather than suppressed growth in a large part of the observable redshift range. In the latter two cases, the phenomenological model gives a poor approximation of the modifications to gravity from an EFTDE, when both are parametrised with the $\propto \Omega_{DE}$ time dependence. The stronger enhancement of growth in the EFTDE model with $\mt > 1, \St < 1$ explains why the posterior for the EFTDE parametrisation is much narrower than the one for the phenomenological parametrisation.

\section{Theory prior II: Different phenomenological time dependences}\label{sec-scale}

The $\propto \Omega_{DE}$ time dependence discussed in \cref{subsec-Sigmu,subsec-eftde} is the most common time dependence used in cosmological analyses of LSS surveys. Naturally, there are several potential pitfalls of using such a highly specific form of the time dependence in the absence of a strong theoretical or observational motivation: for instance, one could miss genuine modified gravity signatures \cite{Linder:2016wqw}, moreover, such a time dependence may be difficult to achieve from any underlying field theoretic model \cite{Linder:2015rcz}. In the absence of a strong motivation for a particular time dependence, we can compare different phenomenological time dependences to distinguish the aspects of modified gravity phenomenology and constraints that depend on the time dependence from those that don't.

With this goal, we compare our results for the $\propto \Omega_{DE}$ time dependence in \cref{sec-mapping} with another time dependence used by some analyses, namely $\propto a$ \cite{Zumalacarregui:2016pph,Noller:2018wyv,Euclid:2025tpw}, where modified gravity parameters are proportional to the scale factor. This time dependence also has the motivation that DE-related effects should be negligible at early times. The key difference between $\propto a$ and $\propto \Omega_{DE}$ is that the $\propto a$ time dependence affects a larger range of redshifts, since $\Omega_{DE}$ decays faster with redshift than $a$. With this time dependence, the phenomenological $\{\mu, \gamma\}$ parametrisation behaves as \cref{mu_scale,gamma_scale}, and the EFTDE $\{\alpha_B, \alpha_M\}$ parametrisation behaves as \cref{alphaB_scale,alphaM_scale}

We now show the analogs of \cref{musigma_signs_omegaDE,mapping_omegaDE} for $\alpha_B, \alpha_M \propto a$ and comment on the qualitative similarities and differences seen for this parametrisation compared to $\alpha_B, \alpha_M \propto \Omega_{DE}$.

\begin{figure}[t]
    \centering
    \includegraphics[width=\linewidth]{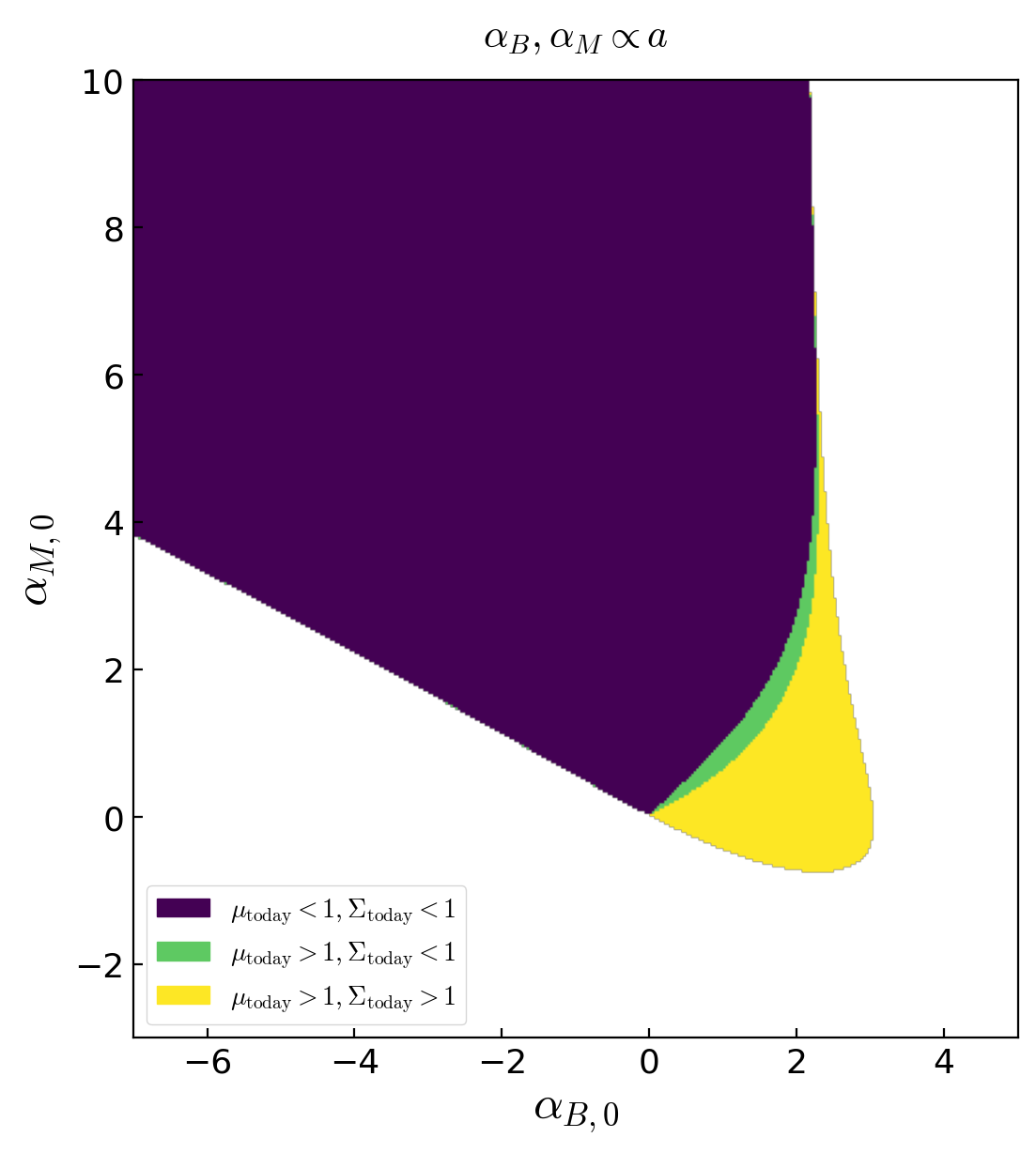}
    \caption{The allowed parameter space for $\BMz$ for the EFTDE parametrisation $\alpha_B, \alpha_M \propto a$. The coloured (white) regions correspond to $\BMz$ values where the gradient stability condition $c_s^2 > 0$ is obeyed (violated) today. The bounds at $\alpha_{M,0}=10$ and $\alpha_{B,0}=-7$ are imposed by hand to limit the size of the region, while the lower bound on $\alpha_{M,0}$ and upper bound on $\alpha_{B,0}$ are physical boundaries imposed by the gradient stability condition. The colours indicate the quadrant of $\mSt$ space that a particular $\BMz$ maps to from evaluating \cref{mu_qsa,Sigma_qsa} at $a=1$. Note the different prior volumes of the quadrants: in particular, the $\mt > 1, \St < 1$ quadrant has the smallest prior volume, in accordance with the conjecture of \cite{Pogosian:2016ji}. The image of this mapping in the $\mSt$ space is shown in \cref{mapping_scale}.}
    \label{musigma_signs_scale}
\end{figure}

First, we compare the two time dependences in the $\BMz$ space by comparing \cref{musigma_signs_scale} with \cref{musigma_signs_omegaDE}. One qualitative difference in \cref{musigma_signs_scale} compared to \cref{musigma_signs_omegaDE} is that the region on the lower right corresponding to $\mt > 1, \St < 1$ in \cref{musigma_signs_omegaDE} is missing in \cref{musigma_signs_scale}. Apart from this, the parameter spaces for the two time dependences are qualitatively similar. For the $\propto a$ time dependence, the $\mt > 1, \St < 1$ quadrant again occupies the lowest prior volume. The $\mt < 1, \St > 1$ quadrant is again absent from the stable parameter space, and the discussion about this in \cref{sec-mapping} for the $\propto \Omega_{DE}$ time dependence directly applies to the $\propto a$ time dependence as well. It is also interesting to compare the observational constraints on $\BMz$ between the two time dependences, which can be done from comparing the right panels of \cref{omegaDE_theory_observations,scale_theory_observations} and the values of constraints from \cref{tab:parameter_constraints}. The data allows for a higher $\alpha_{B,0}$ and a smaller $\alpha_{M,0}$ for the $\propto a$ time dependence compared to the $\propto \Omega_{DE}$ time dependence, and it also leads to a different observational degeneracy between $\BMz$ across the two time dependences.

\begin{figure}[t]
    \centering
    \includegraphics[width=\linewidth]{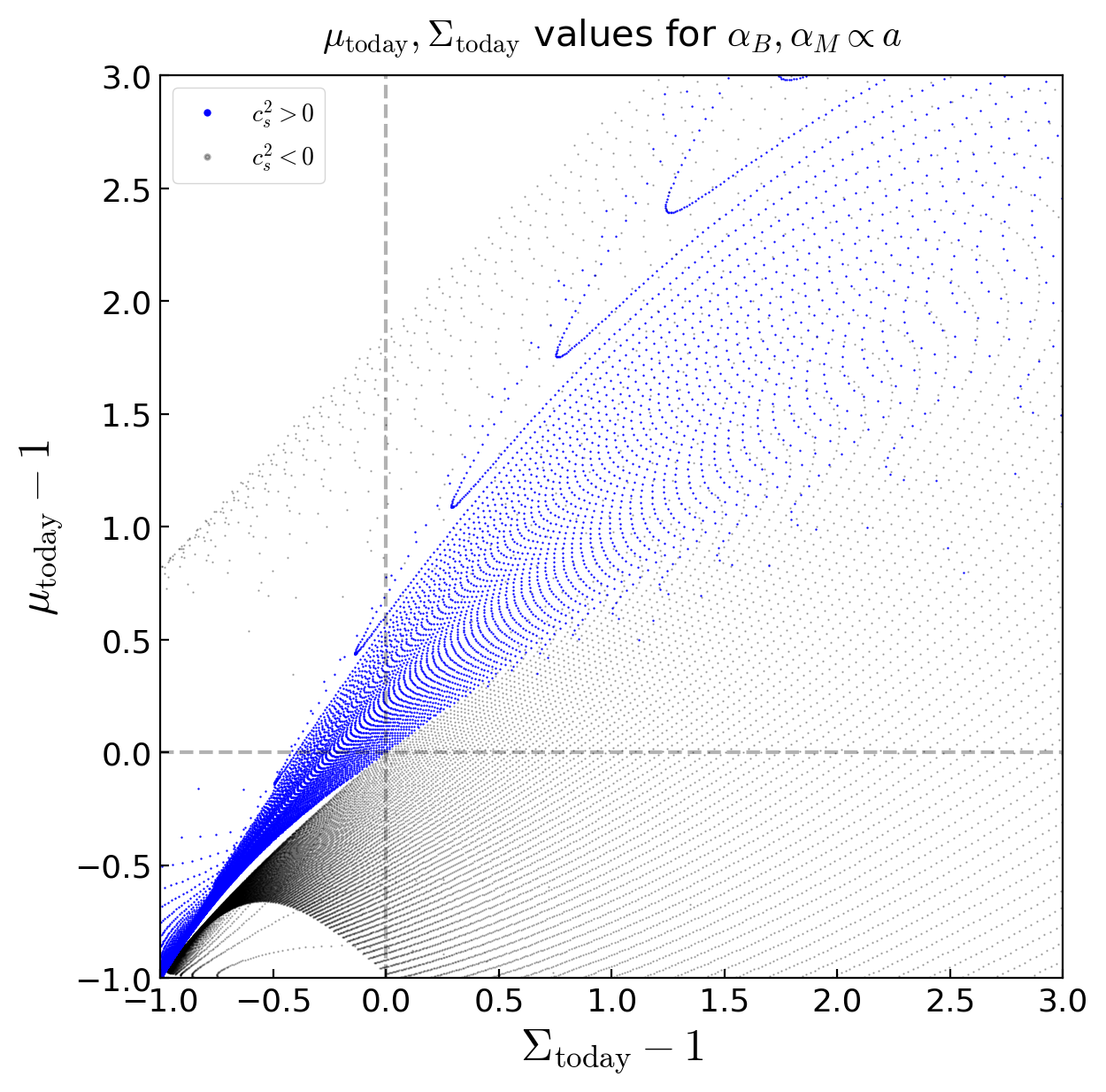}
    \caption{The image of the theoretical parameter space in \cref{musigma_signs_scale} in the $\mSt$ space under the mapping \cref{mu_qsa,Sigma_qsa}. The points shown here are images of a uniform sampling of the $\BMz$ parameter space enclosed in \cref{musigma_signs_scale}. The axis ranges are limited for clarity. Blue (grey) points are derived from $\alpha_{B,0}, \alpha_{M,0}$ values where the gradient stability condition $c_s^2 > 0$ is obeyed (violated) today, which corresponds to the coloured (white) regions in \cref{musigma_signs_omegaDE}. As opposed to \cref{musigma_signs_omegaDE} for the $\propto \Omega_{DE}$ time dependence, here for the $\propto a$ time dependence we find a substantial overlap between the stable and unstable $\mSt$ parameter spaces. \togglered{This feature is caused by a property of the mapping \cref{mu_qsa,Sigma_qsa} that is peculiar to the $\propto a$ time dependence, namely that near-saturation of the gradient stability condition, $c_s^2 \simeq 0$, also numerically corresponds to $\alpha_{B,0} + 2\alpha_{M,0} \simeq 0$, which removes the discontinuity in \cref{mu_qsa,Sigma_qsa} at $c_s^2 \simeq 0$.}}
    \label{mapping_scale}
\end{figure}

\begin{figure*}
    \centering
    \begin{subfigure}
        \centering
        \includegraphics[width=0.48\linewidth]{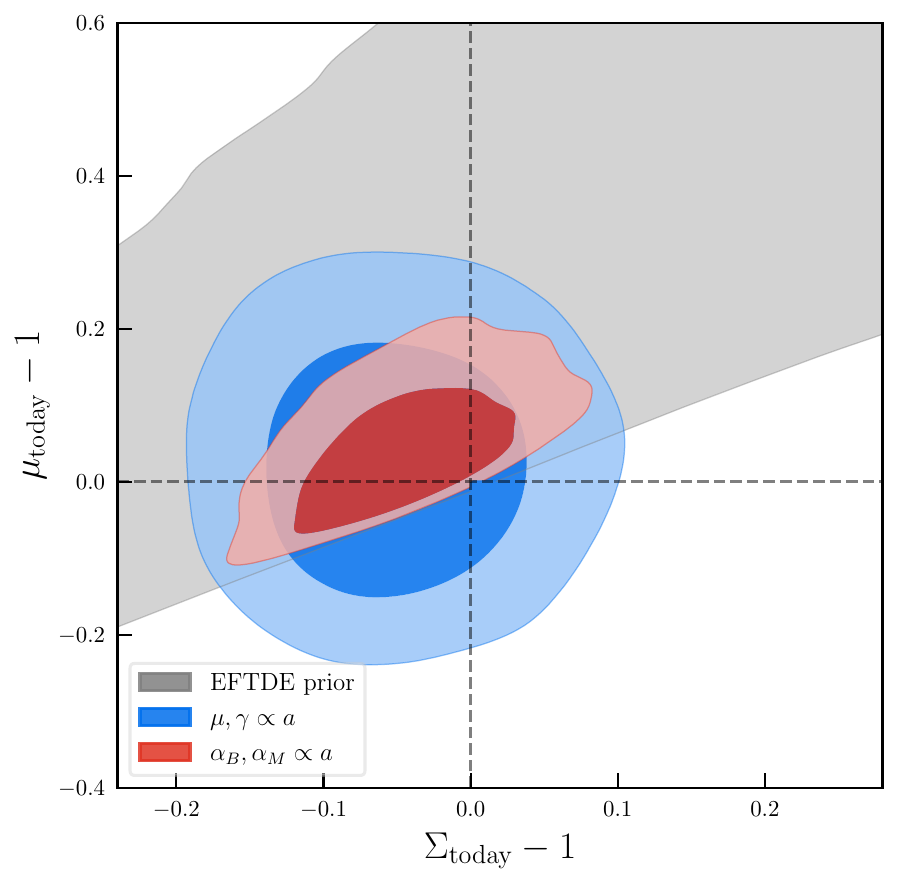}
    \end{subfigure}
    \begin{subfigure}
        \centering
        \includegraphics[width=0.49\linewidth]{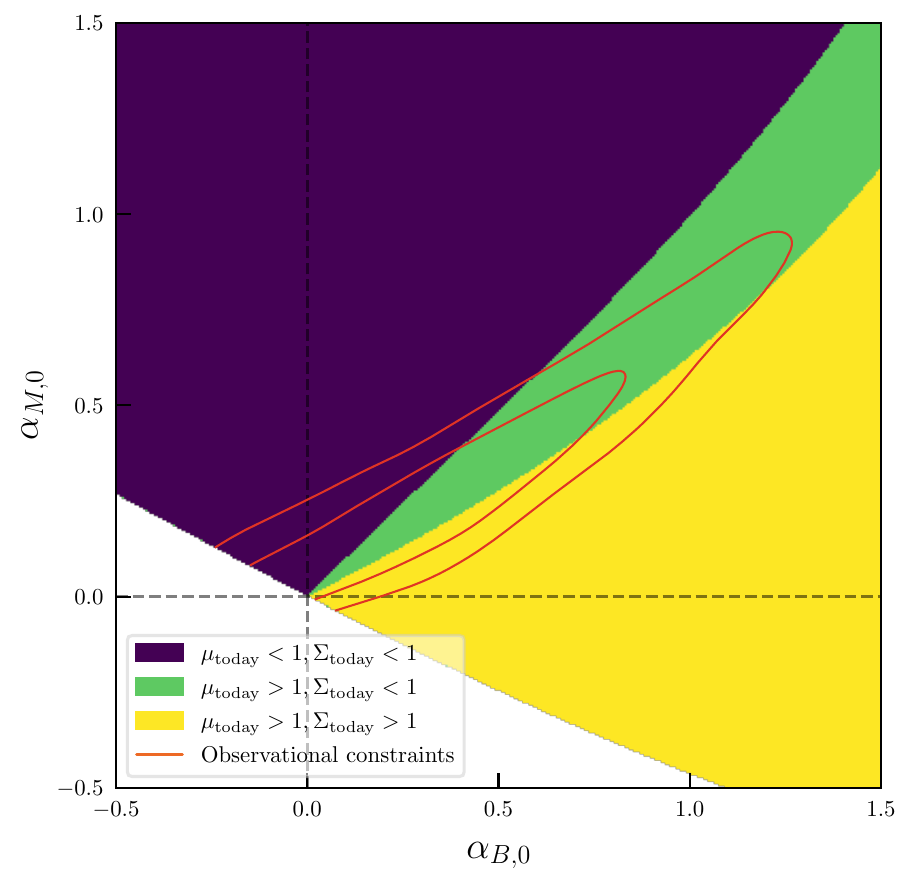}
    \end{subfigure}
    \caption{\textbf{Left panel:} Comparison of observational constraints on $\mSt$ from phenomenological and EFTDE parametrisations for the $\propto a$ time dependence. The grey contour shows the allowed parameter space of the EFTDE prior, with its boundaries arising from the mapping \cref{mu_qsa,Sigma_qsa}. We only show the prior boundaries and not the density of points in the prior, as the relative density of different regions in the prior depends on the prior on the EFTDE parameters as explained in \cref{sec-mapping}. Working with a theory-informed EFTDE parameter basis clearly results in significantly tighter bounds, which is the combined consequence of the EFTDE prior shown here, and the different time evolution of $\{\mu, \Sigma\}$ shown in \cref{musigma_evolution_scale}. \textbf{Right panel:} Comparison of observational constraints with the theoretically allowed parameter space for the $\alpha_B, \alpha_M \propto \Omega_{DE}$ model. An interesting feature of this posterior is that the $\mt > 1, \St < 1$ region which has the smallest relative volume in the prior space has the largest relative volume in the posterior.}
    \label{scale_theory_observations}
\end{figure*}

On the other hand, in the $\mSt$ space, the comparison of the two parameter spaces is more nontrivial. In \cref{mapping_scale}, if we focus only on the stable parameter space with $c_s^2 > 0$, this looks similar both qualitatively and quantitatively to the stable parameter space in \cref{mapping_omegaDE}. Thus, the theoretical prior bounds for the $\propto a$ time dependence would be similar to the ones for the $\propto \Omega_{DE}$ time dependence seen in \cref{omegaDE_theory_observations}. This is shown in \cref{scale_theory_observations}. The roughness at the upper theoretical prior boundary is again due to the sparsity of the mapping \cref{mu_qsa,Sigma_qsa} in that region. Because the mapping for the $\propto a$ time dependence is even more sparse than for the $\propto \Omega_{DE}$ time dependence, we sample a smaller section of the prior volume shown in \cref{musigma_signs_scale}: namely we sample upto $\alpha_{M,0} \leq 6$ instead of $\alpha_{M,0} \leq 10$ in order to get enough samples at the upper boundary of the $\mt > 1, \St < 1$ quadrant in the left panel of \cref{scale_theory_observations}.

The key difference between the two time dependences in the $\mSt$ space is related to the proximity of the stable and unstable parameter spaces. For the $\propto a$ time dependence, there is a high degree of overlap between the stable and unstable $\mSt$ parameter spaces, unlike for the $\propto \Omega_{DE}$ time dependence where they are very separated. To understand this, we take a closer look at the parameter space near the gradient stability divide, which is the parameter space where $c_s^2 \simeq 0$. In the $\BMz$ space, the region with $c_s^2 = 0$ is a smooth gradient stability divide which divides the space into a $c_s^2 > 0$ region and a $c_s^2 < 0$ region. However, the $c_s^2 = 0$ divide has no image in the $\mSt$ space, because the mapping \cref{mu_qsa,Sigma_qsa} contains $c_s^2$ in the denominator. Thus, the values of $\{\mu, \Sigma\}$ are undefined when the gradient stability condition is saturated. Due to the $c_s^2$ in the denominator, one would also have the following two expectations from the $c_s^2 \simeq 0$ (but $\neq 0$) parameter space in the $\mSt$ space: 1) The image in the $\mSt$ space should be far from the GR limit owing to a small magnitude of the $c_s^2$, and 2) The stable and unstable regions around $c_s^2 \simeq 0$ should be widely separated, as they involve a zero crossing in the denominators in both \cref{mu_qsa,Sigma_qsa}.

On looking at \cref{mapping_scale}, it is apparent that expectation 2) fails for the $\alpha_B, \alpha_M \propto a$ time dependence. The form of \cref{mu_qsa,Sigma_qsa} suggests a sufficient condition for it to fail: if the numerators of both \cref{mu_qsa,Sigma_qsa} undergo a zero crossing at the same time as the denominator in the $\alpha_{B,0}, \alpha_{M,0}$ parameter space. This can occur only if $c_s^2 \simeq 0$ corresponds to $\alpha_B + 2\alpha_M = 0$ in the $\BMz$ parameter space. This is easy to check numerically from \cref{cs2}, and indeed we find that this holds to a good approximation (specifically, $\alpha_{M,0}/\alpha_{B,0} = -1/2$ within 10\%) around $\alpha_{B,0} \simeq 0$. We can also mathematically derive this property close to the GR limit by linearising \cref{cs2} in $\BMz$. By setting the linearised $c_s^2$ to zero at $z=0$, we find that the gradient stability divide occurs at $\alpha_{M,0}/\alpha_{B,0} \simeq -0.53$ for the $\propto a$ time dependence, while it occurs at $\alpha_{M,0}/\alpha_{B,0} \simeq -0.67$ for the $\propto \Omega_{DE}$ time dependence, for which we see that the stable and unstable regions in the $\mSt$ space are indeed separated.

\begin{figure*}
    \centering
    \begin{subfigure}
        \centering
        \includegraphics[width=0.49\linewidth]{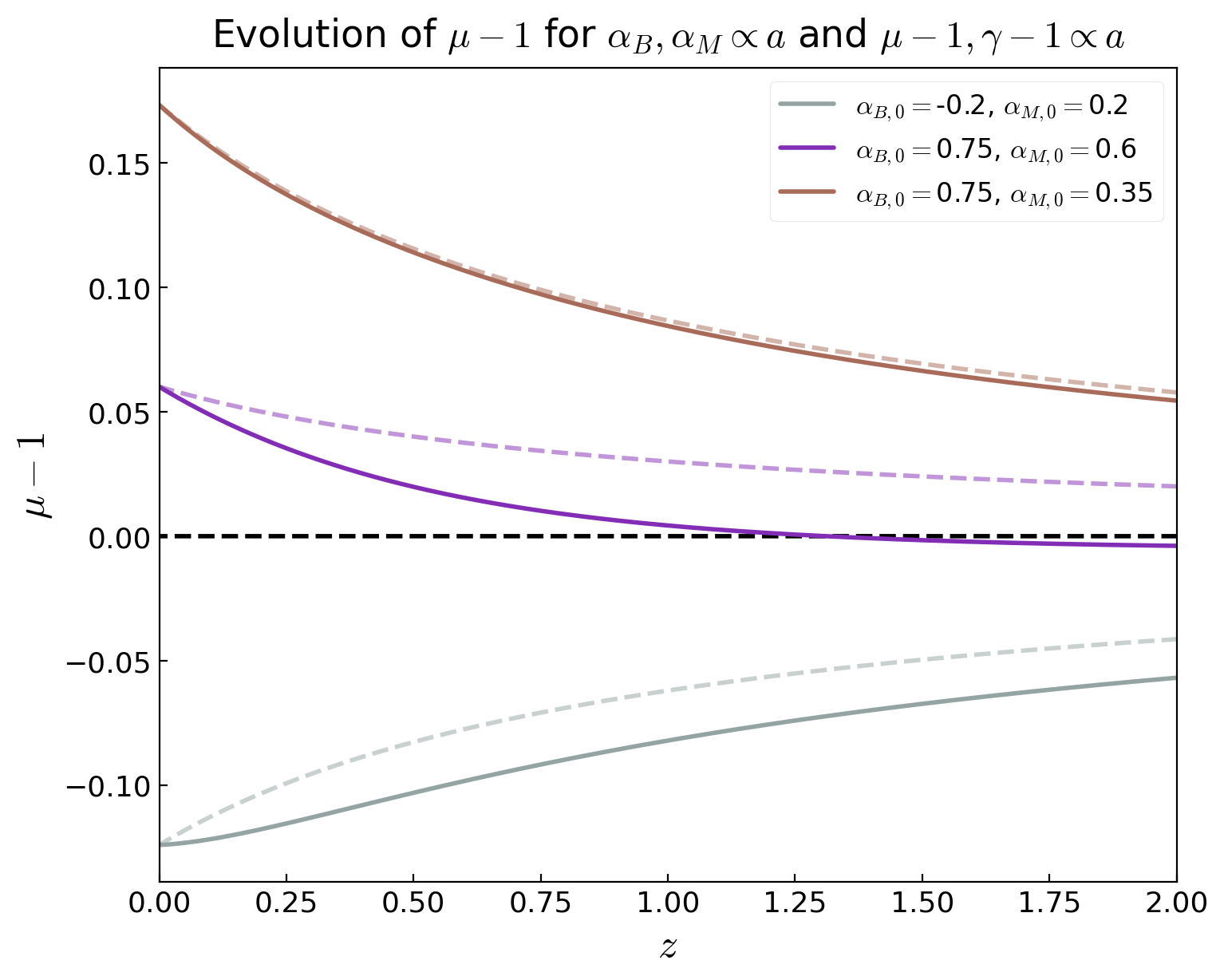}
    \end{subfigure}
    \begin{subfigure}
        \centering
        \includegraphics[width=0.48\linewidth]{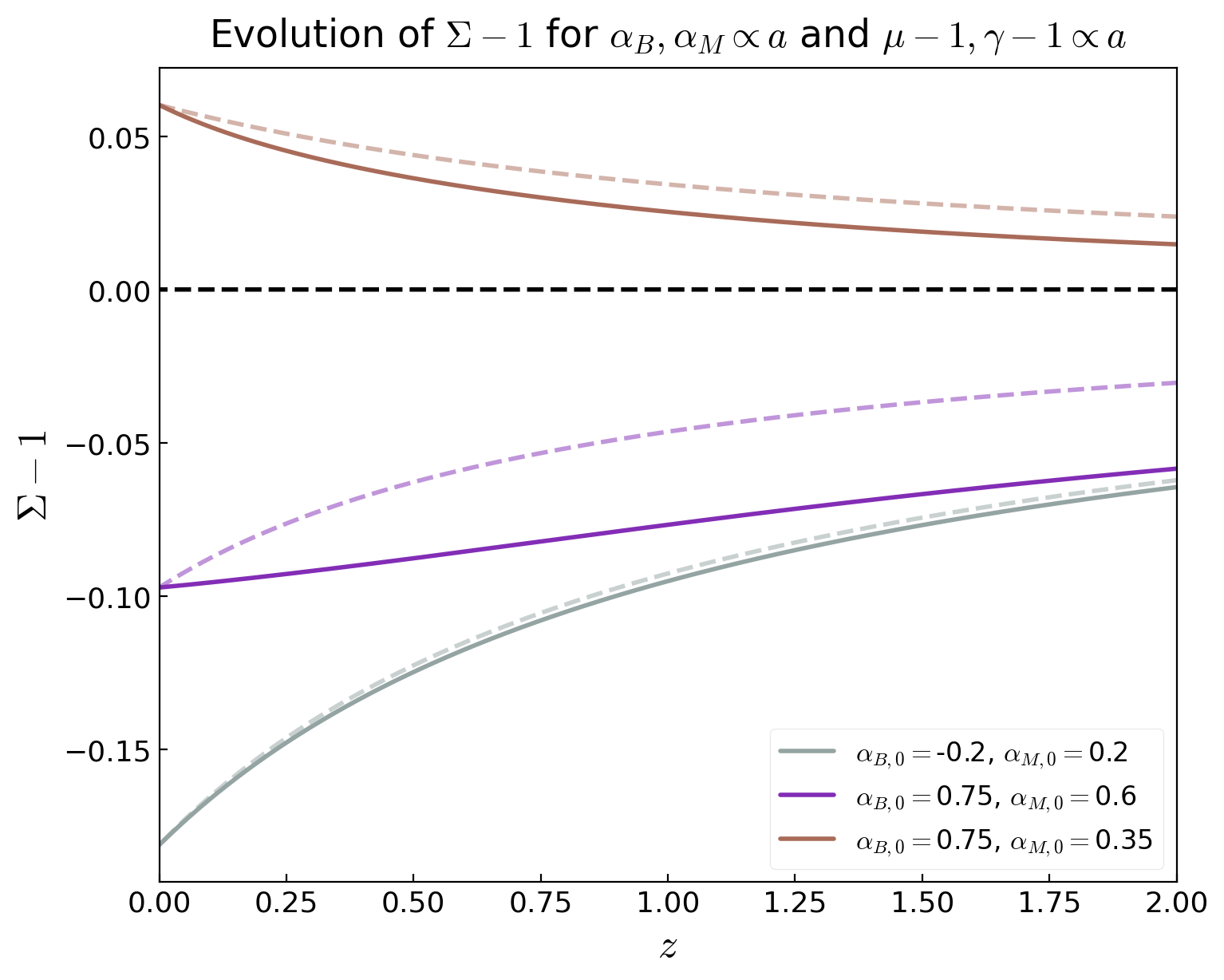}
        
    \end{subfigure}
    \caption{The time evolution of $\mu$ and $\Sigma$ for EFTDE and phenomenological parametrisations for the $\propto a$ time dependence. The solid lines are the time evolution in the EFTDE parametrisations, while the faint dashed lines are the time evolution in the phenomenological parametrisations with the same $\mSt$ values. We show the time evolutions for one point in each of the three observationally allowed quadrants in the $\mSt$ space for EFTDE parametrisations (see \cref{sec-mapping}). Compared to the $\propto \Omega_{DE}$ time dependence in \cref{musigma_evolution_omegaDE}, the phenomenological and EFTDE parametrisations match more closely for the $\propto a$ time dependence.}
    \label{musigma_evolution_scale}
\end{figure*}

\begin{figure}[htp]
    \centering
    \includegraphics[width=\linewidth]{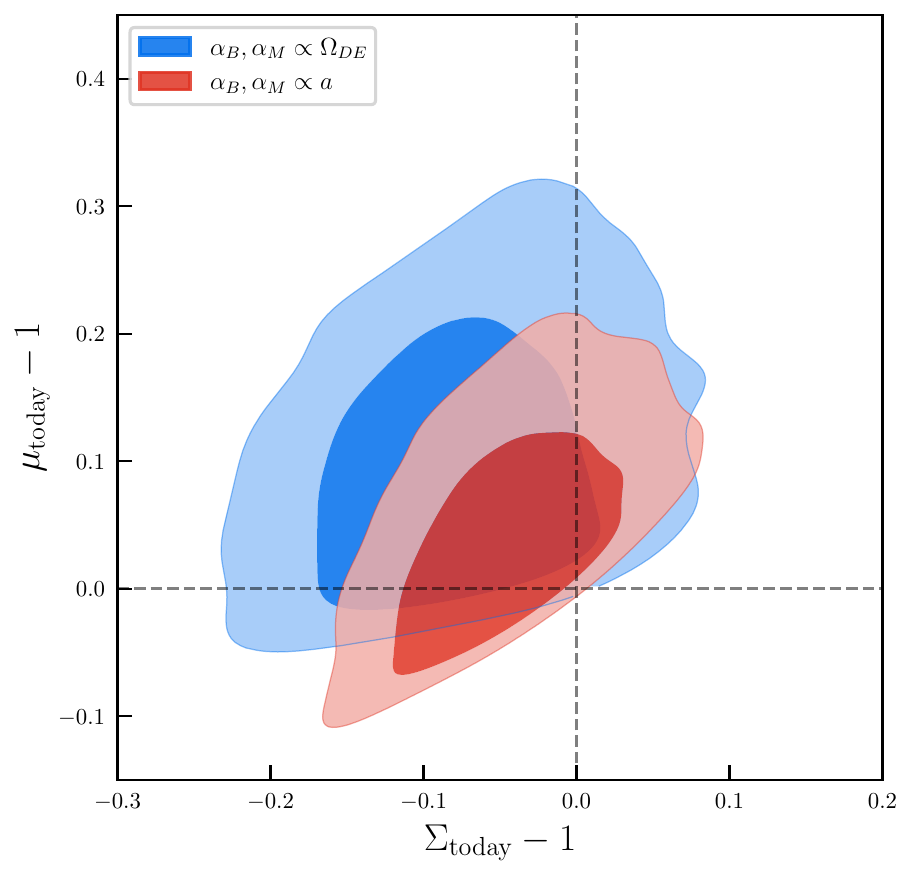}
    \caption{Comparison of constraints on $\mSt$ for the EFTDE parametrisation between the $\propto \Omega_{DE}$ and $\propto a$ time dependences. The $\propto a$ time dependence has tighter posteriors for $\mSt$ compared to the $\propto \Omega_{DE}$ time dependence.}
    \label{omegaDE_scale_eftde}
\end{figure}

\begin{figure}[htp]
    \centering
    \includegraphics[width=\linewidth]{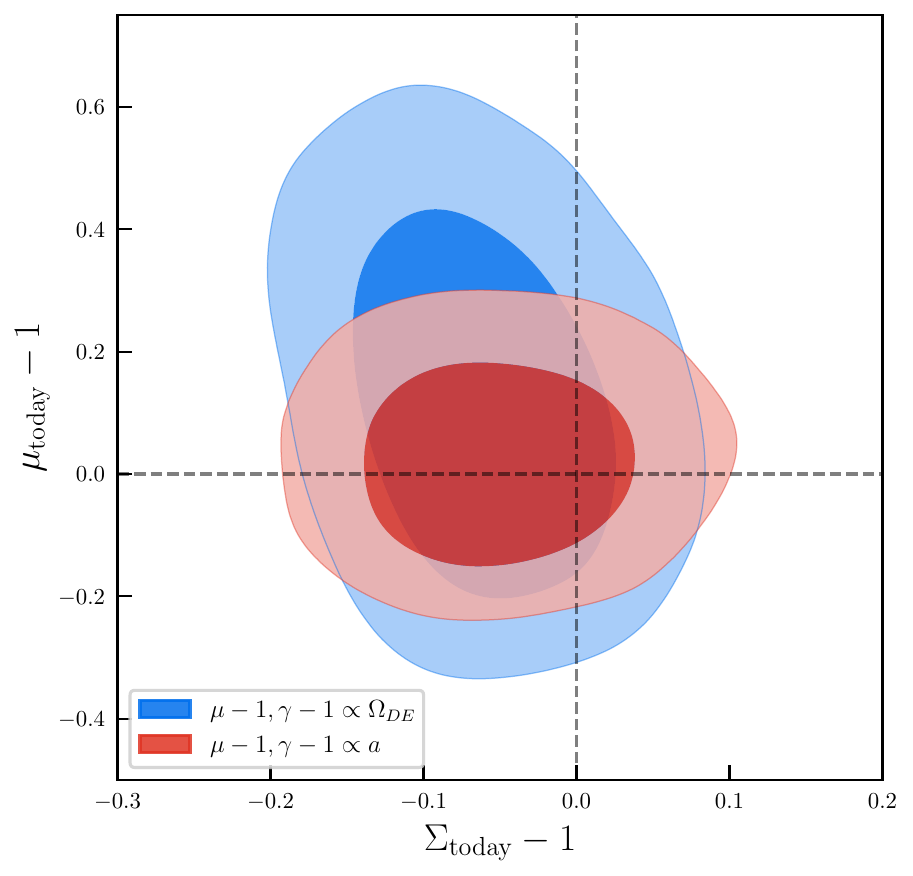}
    \caption{Comparison of constraints on $\mSt$ for the phenomenological parametrisation between the $\propto \Omega_{DE}$ and $\propto a$ time dependences. The $\propto a$ time dependence has a significantly tighter posterior for $\mt$ compared to the $\propto \Omega_{DE}$ time dependence. However, the choice of the time dependence has little effect on the posterior for $\St$.}
    \label{omegaDE_scale_Sigmu}
\end{figure}

We now compare the $\propto \Omega_{DE}$ and $\propto a$ time dependences at the level of the posteriors of $\mSt$. We do this for the EFTDE parametrisations in \cref{omegaDE_scale_eftde}, and for the phenomenological parametrisations in \cref{omegaDE_scale_Sigmu}. For both the EFTDE and phenomenological parametrisations, we find tighter constraints on both $\mt$ for the $\propto a$ time dependence, while the constraints on $\St$ are tightened for the EFTDE parametrisation but not for the phenomenological parametrisation. The impact on $\St$ constraints is smaller because the observational bounds on $\St$ are fairly tighter than the constraints on $\mt$ for both time dependences. The constraints on $\mt$ are tighter for the $\propto a$ time dependence because the effects of modified gravity decay more slowly with redshift for this time dependence, and so our higher redshift data has more constraining power than it does for the $\propto \Omega_{DE}$ time dependence. The slower decay with redshift of $\{\mu, \Sigma\}$ for both the phenomenological and EFTDE parametrisations for the $\propto a$ time dependence can be seen in \cref{musigma_evolution_scale}. \cref{musigma_evolution_scale} also shows that for the $\propto a$ time dependence, the phenomenological parametrisation does a relatively better job at approximating the modified gravity phenomenology of the EFTDE parametrisation, compared to the $\propto \Omega_{DE}$ time dependence in \cref{musigma_evolution_omegaDE}.

\section{Theory prior III: Theory-informed time dependence}\label{sec-shiftsym}

In \cref{sec-Param}, we introduced the same time dependence $\mu-1, \gamma-1 \propto \Omega_{DE}$ and $\alpha_i \propto \Omega_{DE}$ for the phenomenological and theory-informed parametrisations of modified gravity. Apart from the requirement for modified gravity effects to be contemporaneous with dark energy domination, there is no other physical motivation for this particular time dependence. The assumption of this time dependence thus selects a highly biased subset of all modified gravity theories (both phenomenological and within EFTDE) without any a-priori theoretical motivation for this bias. In the absence of a strong theoretical or observational preference for a particular time dependence, the only way in which we can mitigate this bias is by comparison with a different but equally ad-hoc time dependence -- this is what we do in \cref{sec-scale}.

However, for special classes of EFTDE, it is possible to impose a theoretical prior on the space of all possible time dependences (or at least a significant subset of it). This exercise has been done for the class of shift-symmetric theories with a trivial speed of GWs \cite{Traykova:2021hbr} and results in a prior characterising the time dependence of the underlying EFT function(s). Namely, \cite{Traykova:2021hbr} proposed the following fit for the time dependence of $\alpha_B$ in the context of the aforementioned theories 
\begin{gather}\label{shiftsym}
\alpha_B = \alpha_{B,0}\left(\dfrac{H}{H_0}\right)^{-\frac{4}{m}}\ ,
\end{gather}
which covers a wide range of monotonic time dependences of $\alpha_B$ where $\alpha_B$ vanishes at early times. 

The condition of luminal GW speed, setting $\alpha_T = 0$, combined with the condition of shift symmetry, restricts $\alpha_M$ to $\alpha_M = 0$. As we will discuss in \cref{sec-GWpriors}, the prior $\alpha_T = 0$ is not trivial to motivate just from the observation of a luminal GW speed at the frequencies probed by current GW detectors. Thus, while we set $\alpha_M = \alpha_T = 0$ in this section, we highlight that this is a model choice and not enforced by some overarching physicality prior.

\begin{figure}[t]
    \centering
    \includegraphics[width=\linewidth]{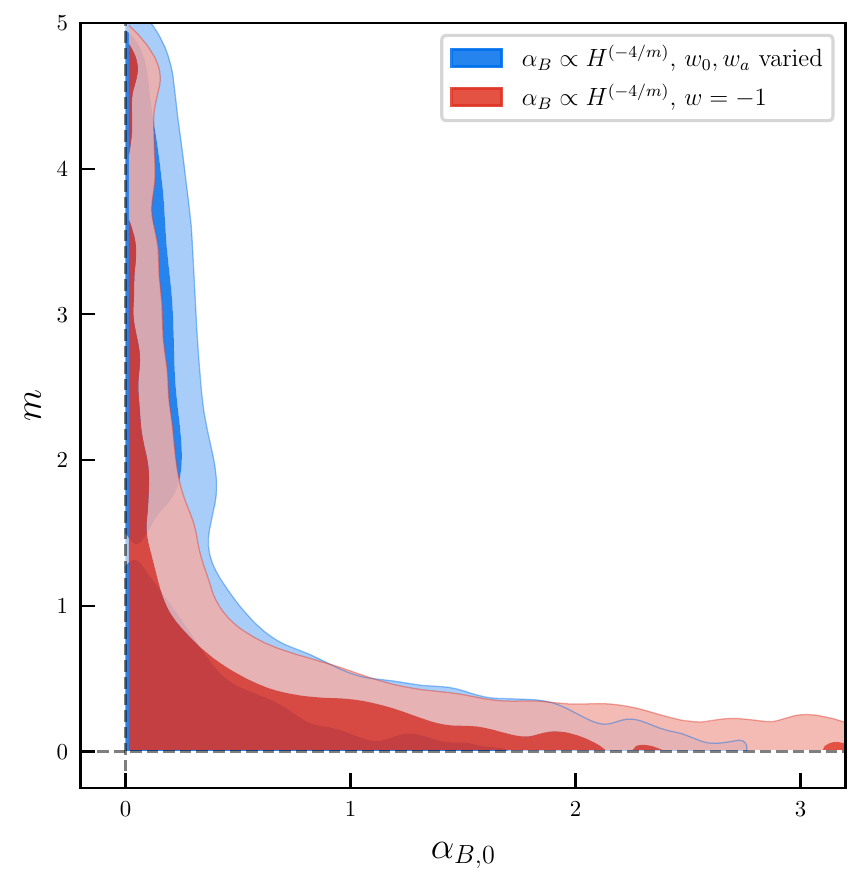}
    \caption{Constraints on the EFTDE parameters $\alpha_{B,0}$ and $m$ for shift symmetric theories with the theoretically motivated time dependence \cref{shiftsym}, for the $\Lambda$CDM and CPL backgrounds. The constraints are consistent with the GR limit, which for this parametrisation is given by any of the two entire lines $\alpha_{B,0} = 0$ and $m=0$, not just the point where both $\alpha_{B,0}$ and $m$ are zero.}
    \label{hpl}
\end{figure}

From the expressions \cref{mu_qsa,Sigma_qsa}, we see that these theories have $\mu(a) = \Sigma(a) > 1$. In \cite{Traykova:2021hbr}, it was also shown that when the background expansion history is allowed to vary, the CPL parametrisation \cref{w0wa} is sufficiently general to capture the relevant theory space. We show the constraints on the EFTDE parameters $\alpha_{B,0}, m$ in \cref{hpl}, for both the $\Lambda$CDM and CPL backgrounds. Our constraints differ from the constraints found by \cite{Traykova:2021hbr} on the same model due to the difference in datasets.

For this parametrisation, the GR limit is not at a single point but on both of the lines $\alpha_{B,0} = 0$ and $m = 0$. The $m=0$ case is the GR limit only in the phenomenological sense, because it corresponds to the case where modified gravity is turned off at all observable redshifts in the past but turns on precisely at $z=0$ where we don't have any cosmological observations. Due to this, the posteriors for both $\alpha_{B,0}$ and $m$ have long tails if they are consistent with GR.

\begin{figure}[t]
    \centering
    \includegraphics[width=\linewidth]{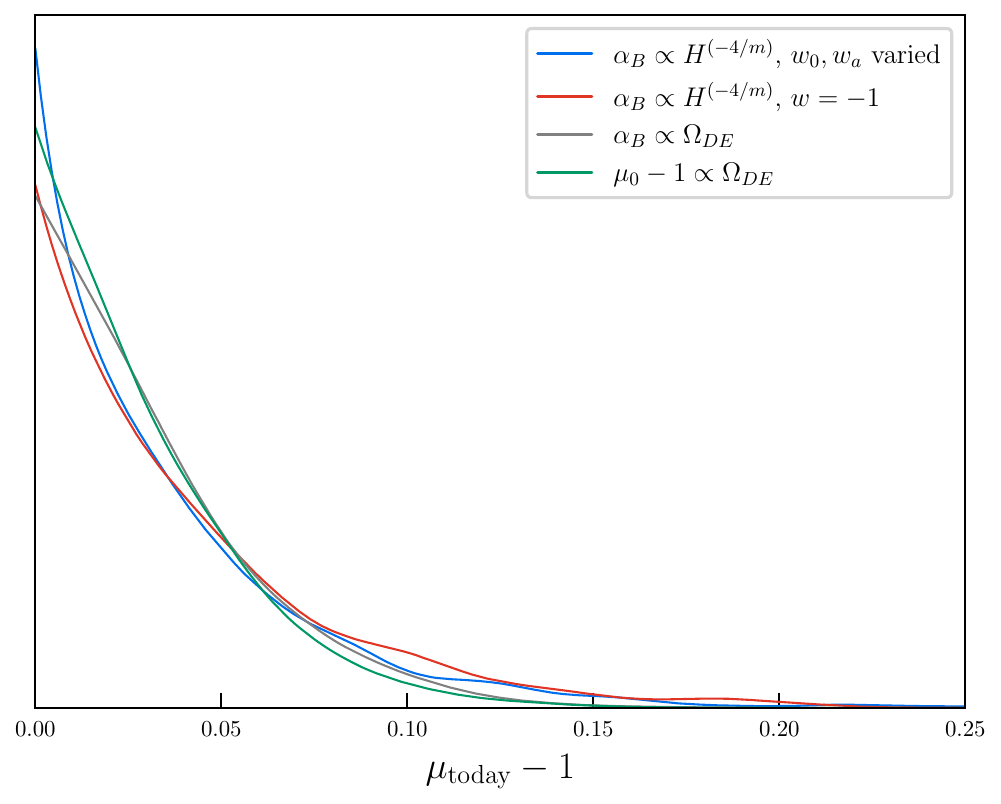}
    \caption{A comparison of the posteriors of $\mt - 1 = \St - 1$ across various phenomenological and theory-informed models of no-slip gravity (no-slip is equivalent to the condition $\mu(a) = \Sigma(a)$). We find that all these models with varying degrees of theoretical motivations give very similar posteriors on $\mt$, and thus cannot be distinguished from the constraints on $\mt$.}
    \label{noslip_comparison}
\end{figure}

From the viewpoint of observational constraints, it is also important to ask whether the more theoretically motivated parametrisations of time dependence can actually be distinguished from naive phenomenological ones -- in many cases they can't \cite{Gleyzes:2017kpi}. We now answer this question for the time dependence \cref{shiftsym}, by comparing its observational posterior on $\mt$ with more naive parametrisations that lead to a similar phenomenology of modifications to gravity. In particular, in these shift symmetric theories, we have $\mu(a) = \Sigma(a) > 0$. This translates to the two metric potentials $\Phi$ and $\Psi$ being equal, a condition referred to as "no-slip". In \cref{noslip_comparison}, we show the 1D posterior of $\mu$ for the shift symmetric parametrisation and compare it with two naive parametrisations that produce no-slip: I) $\alpha_B \propto \Omega_{DE},\ \alpha_M = \alpha_T = 0$, where the choice of dynamical DE functions is theoretically motivated but the time dependence isn't, and II) $\mu > 1, \mu \propto \Omega_{DE}, \gamma = 1$ where both the dynamical DE functions and their time dependence are phenomenologically parametrised (in the phenomenological parametrisation, $\mu > 1$ is imposed by hand for a fairer comparison with the EFTDE parametrisations, which exclude $\mu < 1$ from the gradient stability requirement). For the latter two cases we do not show the comparison with varied expansion histories, as we will show in \cref{sec-background} that marginalising over the background expansion history does not affect constraints on $\mSt$ for the $\propto \Omega_{DE}$ time dependence. All of these 1D posteriors are much narrower than the 1D posteriors for $\mt$ or $\St$ obtained by marginalising the 2D posteriors in more general phenomenological or EFTDE parametrisations (as seen from \cref{omegaDE_theory_observations} and \cref{scale_theory_observations}). This can be understood from the fact that the posteriors for the more general theories prefer enhanced growth and reduced lensing while the no-slip condition forces the effects on growth and lensing to have the same sign. \cref{noslip_comparison} also suggests that we cannot meaningfully distinguish any of the various no-slip models from the parameter constraints. We conclude that for shift symmetric theories the phenomenological parametrisation is sufficient to capture the effects of EFTDE theories in terms of the constraints on $\mt$.

\section{Theory prior IV: Gravitational wave priors}\label{sec-GWpriors}

The theoretical priors discussed in the sections above all restrict the functional dependence of the theory functions controlling the evolution of cosmological perturbations.\footnote{In some cases this is the case for subsets of theories, where certain theory functions are assumed to vanish (cf. section \ref{sec-shiftsym}), but the prior itself does not provide any strong rationale for eliminating these functions itself.} Unlike the above priors, an even more powerful prior would completely eliminate (some combination of) functions in the $\alpha_i$ or $\{\mu,\Sigma\}$ bases, thereby reducing the dimensionality of the space of functions to be constrained. Priors derived from considering the propagation of gravitational waves do so by providing complementary constraints to those derived from large scale structure evolution alone and can therefore be highly restrictive. The two priors we will focus on are 1) a prior on the speed of gravitational waves, enforcing this to be luminal and hence setting $\alpha_T = 0$, and 2) a prior from considering the stability of dark energy perturbations on backgrounds sourced by `loud' gravitational wave mergers. 

A prior setting $\alpha_T =0$ is frequently motivated by the near-coincident observation of the binary neutron star merger GW170817 in gravitational waves and optical follow-up surveys \cite{TheLIGOScientific:2017qsa,Goldstein:2017mmi,Savchenko:2017ffs,LIGOScientific:2017zic,GBM:2017lvd}. However, the fact that this observation takes place at frequencies ${\cal O}(10 - 10^3)$ Hz, while the naive cutoff of large classes of cosmologically motivated scalar-tensor theories (in fact, in particular all of the theories we consider here which give $\alpha_T \neq 0$ on cosmological scales) is $\Lambda_3 \sim {\cal O}(10^2)$ Hz \cite{deRham:2018red}, means it is highly non-trivial to map this constraint to the cosmological regime and this can in fact only be done rigorously by imposing additional assumptions on the frequency-dependence of $\alpha_T$ - also see related discussions in \cite{Harry:2022zey,LISACosmologyWorkingGroup:2022wjo,Baker:2022eiz,Sirera:2023pbs,Atkins:2024nvl}. As such, while we impose the simplifying $\alpha_T = 0$ throughout most analyses presented in this paper, in this section we in addition investigate the more conservative scenario where no additional assumption on $\alpha_T$ is imposed and this function remains free.

Secondly, when $\alpha_T = 0$ is assumed, \cite{Creminelli:2019kjy} showed that dark energy perturbations develop (gradient) instabilities on gravitational wave backgrounds (e.g. sourced by the loud mergers of supermassive black hole binaries) unless one imposes
\begin{gather}
|\alpha_B + \alpha_M| \lesssim {\cal O}(10^{-2}).
\end{gather}
Given the expected constraining power of present-day cosmological analyses this effectively amounts to imposing a $\alpha_M = -\alpha_B$ prior. This prior (together with imposing $\alpha_T = 0$) then effectively reduces an otherwise three-dimensional $\alpha_i$ functional parameter space $\{\alpha_B, \alpha_M, \alpha_T\}$ (where we exclude the observationally sub-dominant $\alpha_K$) to a one-dimensional functional parameter space. Here, we consider the fiducial time dependence $\propto \Omega_{DE}$ for the surviving theory function $\alpha_B$ (or $-\alpha_M$), thus the parametrisation that we study in this section is given by

\begin{gather} \label{GWprior_omegaDE}
\alpha_B = -\alpha_M \propto \Omega_{DE},\ \alpha_T = 0\ .
\end{gather}

\subsection{The speed of gravitational waves}

\begin{figure}[t]
    \centering
    \includegraphics[width=\linewidth]{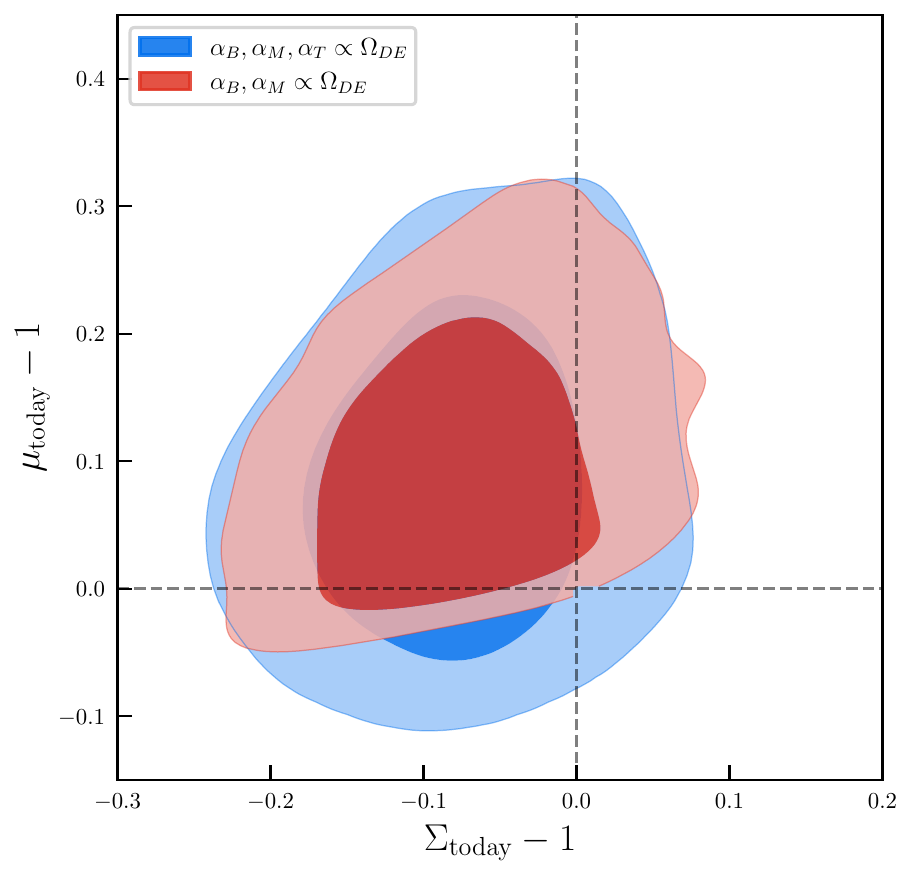}
    \caption{Comparison of constraints on $\mt, \St$when $\alpha_{T,0}$ is marginalised over with the case where it is fixed to zero. For a varying $\alpha_T$, we find a broader posterior in the $\mt < 1, \St < 1$ quadrant, and also some posterior volume in the $\mt < 1, \St > 1$ quadrant which is entirely inaccessible when $\alpha_T = 0$.}
    \label{omegaDE_alphaT_musigma}
\end{figure}

In \cref{omegaDE_alphaT_musigma}, we compare the posteriors on $\mSt$ between the EFTDE model $\alpha_B, \alpha_M \propto \Omega_{DE}, \alpha_T=0$ and the model with $\alpha_B, \alpha_M, \alpha_T \propto \Omega_{DE}$. For the latter model, our results for constraints on $\alpha_{T,0}$ are similar to those of \cite{Noller:2018wyv}. 

We find a slight broadening on the $\mt-\St$ posterior in the $\mt, \St < 1$, and even in the $\mt < 1, \St > 1$ quadrant which is not accessible when $\alpha_{T,0}$ is fixed to zero. Empirically, we have found that these new points in the $\mt < 1, \St > 1$ quadrant have $\alpha_T < 0$, i.e. subluminal GW speeds. It is not straightforward to predict this analytically, as the quasistatic expressions for $\{\mu, \Sigma\}$ when all three of $\{\alpha_B, \alpha_M, \alpha_T\}$ are varied are not very transparent (these expressions can be found in, e.g., \cite{Bellini:2014fua,Gleyzes:2014rba,Linder:2015rcz,Pogosian:2016ji,Noller:2020afd}).

\subsection{Stability in a GW background}

\begin{figure}[t]
    \centering
    \includegraphics[width=\linewidth]{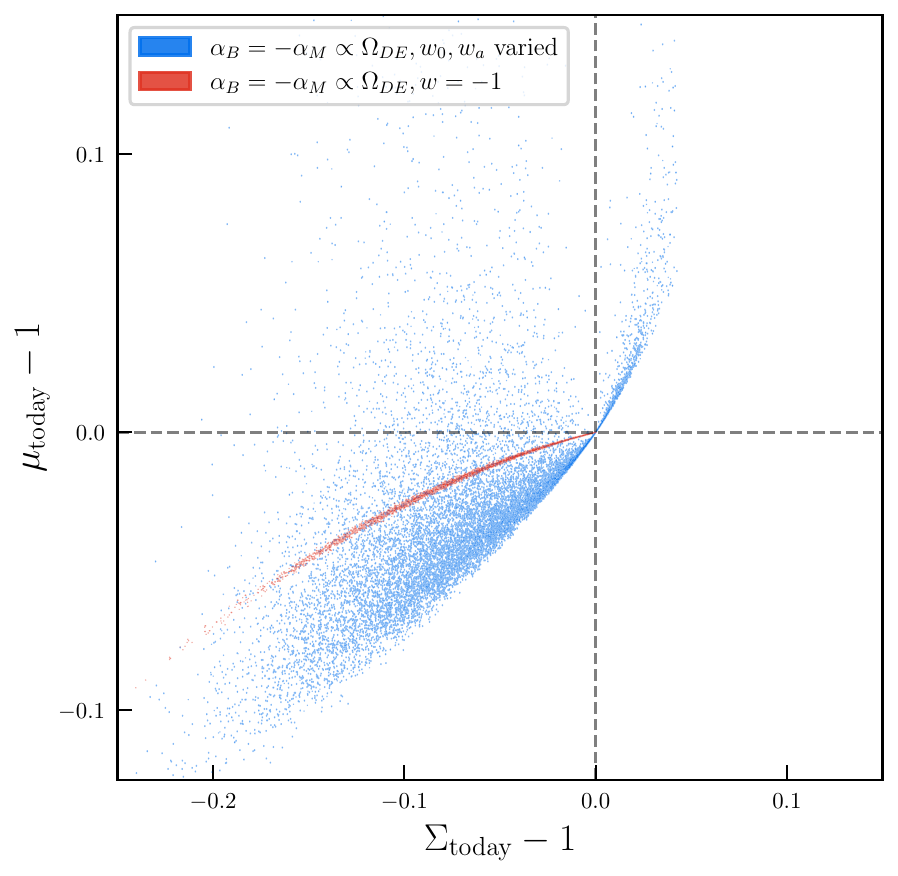}
    \caption{Constraints on $\mSt$ derived from EFTDE after imposing the GW decay prior $\alpha_B + \alpha_M = 0$. We have chosen not to smooth these posteriors due to their unusual shapes. Two key features jump out: 1) The reduction from a two- to a one-dimensional parameter space when $\alpha_B = -\alpha_M$ is imposed for a fixed $w=-1$ background, and 2) the broadening of the resulting contour when the background is varied and marginalised over instead. (Conversely, this parametrisation of perturbations also has a strong effect on the background constraints -- details in \cref{sec-background}.)}
    \label{omegaDE_GW_w0wa}
\end{figure}

We show the observational posteriors for $\mSt$ for the GW-stability motivated parametrisation \cref{GWprior_omegaDE} in \cref{omegaDE_GW_w0wa}. When the background expansion history is fixed to $\Lambda$CDM, these theories have only a single degree of freedom in the model, namely $\alpha_B$. This reflects in the $\mSt$ posterior by making it almost perfectly one-dimensional, as seen from \cref{omegaDE_GW_w0wa}. We have chosen not to smooth the posterior so that its very small but nonzero thickness is accurately represented. Its thickness is due to marginalising over $\Omega_{DE,0}$ (equivalently, $\Omega_{m,0}$), which is tightly constrained by the combination of SNe, BAO, and CMB data. For a $\Lambda$CDM background, the gradient stability condition enforces $\alpha_{B,0} < 0$ and consequently $\alpha_{M,0} > 0$ for this model. The requirement of $\alpha_{M,0} > 0$ explains why the posterior lies entirely in the $\mt > 1, \St > 1$ quadrant, as a high Planck mass suppresses $\mSt$.

When the background expansion history is allowed to vary, there are now two degrees of freedom in the model, which results in a 2D posterior for $\mSt$. This posterior has a very unusual shape as shown in \cref{omegaDE_GW_w0wa}. An interesting feature of this posterior is the bifurcation around the $\St = 1$ line. The reason for this is that in these theories, from \cref{Sigma_qsa}, $\Sigma = 1/M^2$, and thus the only way to achieve $\Sigma = 1$ is to have $\alpha_M = 0 = \alpha_B$, which also enforces $\mu = 1$. Whereas for $\Sigma \neq 1$, there is enough freedom between $\alpha_B$ and $w$ to allow for multiple values of $\mu$ for the same value of $\Sigma$. The freedom in the expansion history also allows the posterior to extend in the $\mt > 1, \St  < 1$ and $\mt > 1, \St > 1$ quadrants. Thus, the impact of the assumed expansion history on these theories is much more substantial than the more general EFTDE and phenomenological models discussed in the previous sections. The converse is also true: this particular parametrisation of the perturbations places a strong theoretical prior on the background expansion history. This is discussed in detail in the next section.

\section{Interplay of constraints on the background and perturbations}\label{sec-background}

\begin{figure}[t]
    \centering
    \includegraphics[width=\linewidth]{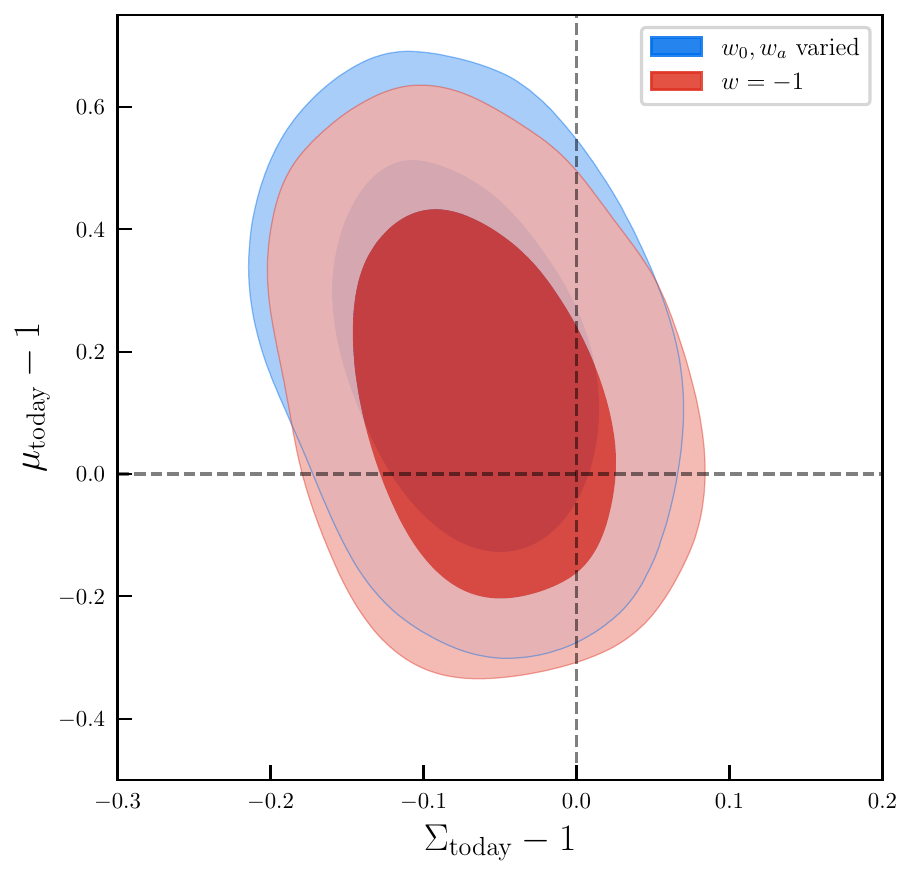}
    \caption{Comparison of constraints on $\mt, \St$when $w_0, w_a$ are marginalised over with the case where $w(a)$ is fixed to $-1$ for the $\mu - 1, \gamma - 1 \propto \Omega_{DE}$ parametrisation.}
    \label{Sigmu_w0wa_comparison_MG}
\end{figure}

Throughout this paper, we have parametrised the background and perturbation dynamics of modified gravity independently. In phenomenological parametrisations, this approach is justified as our probes of background expansion and structure growth are largely independent of each other. This is confirmed by looking at \cref{Sigmu_w0wa_comparison_MG}, which compares the posteriors on $\mSt$ for the phenomenological parametrisation \cref{mu_OmegaDE,gamma_OmegaDE} between the case with a $\Lambda$CDM background and the case with a CPL background parametrised as \cref{w0wa}, and constrained with the CMB, BAO, and supernovae data described in \cref{sec-data}. In the phenomenological parametrisation, there is no theory prior imposed by the background on the perturbations and vice versa, and \cref{Sigmu_w0wa_comparison_MG} comfirms the expectation that freeing up the background expansion history has a negligible effect on the constraints on $\mSt$.

However, in the EFTDE framework, the dynamics of a single field (including its coupling to gravity) both drives the accelerated expansion of the universe and affects the growth of structure. In this case, it is not obvious how the constraints on the modified gravity perturbations depend on the parametrisation of the background expansion, and vice versa. In particular, the enforcement of the gradient stability condition $c_s^2 > 0$, with $c_s^2$ given by \cref{cs2}, introduces a theoretical prior jointly on the background and perturbation degrees of freedom, as the expression for $c_s^2$ involves both the expansion history and the functions parametrising perturbations. We now study how the parametrisation of the background expansion affects constraints on the perturbations and vice versa in EFTDE models.

\begin{figure}[t]
    \centering
    \includegraphics[width=\linewidth]{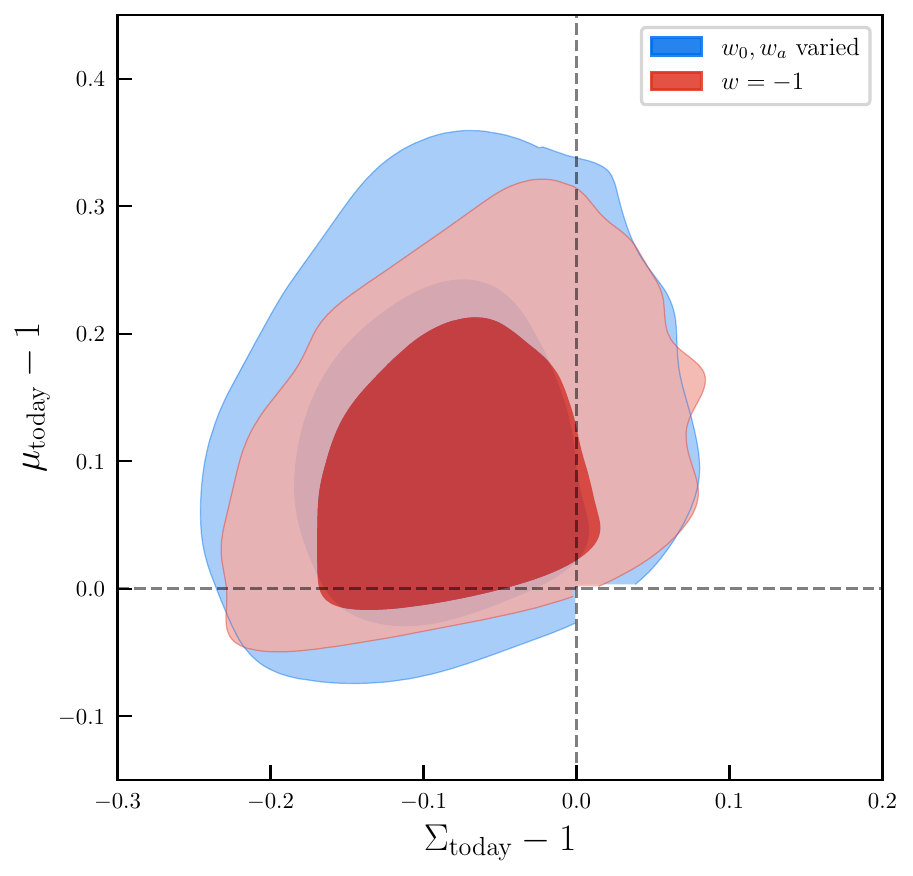}
    \caption{Comparison of constraints on $\mSt$ when $\{w_0, w_a\}$ are marginalised over with the case where $w(a)$ is fixed to $-1$ for the $\alpha_B, \alpha_M \propto \Omega_{DE}$ parametrisation. Note that for the varying $\{w_0, w_a\}$ case, we have cut the $\mu < 1, \St > 1$ quadrant from the posterior as we have found no samples in that quadrant from our chains.}
    \label{omegaDE_w0wa_musigma}
\end{figure}

We first focus on the effect of marginalising over the background expansion history on the constraints on modified gravity perturbations. \cref{omegaDE_w0wa_musigma} compares the posteriors on $\mSt$ derived from the EFTDE parametrisation \cref{eqn:baseline_alphas}, between a $\Lambda$CDM background and a CPL background \cref{w0wa}. Similarly to the phenomenological parametrisation, we only find a slight broadening of the $\mSt$ posterior for the EFTDE parametrisation. This indicates that the values of $\mSt$ are more sensitive to the values of the perturbation functions than to the background expansion. For the CPL background, there is still no posterior volume in the $\mt < 1, \St > 1$ quadrant (we have checked this directly from the samples in our chains, and then cut this quadrant from the posterior in \cref{omegaDE_w0wa_musigma} to avoid a smoothing artefact in that contour). This is because the mathematical condition for this to occur, \cref{musigma_fourthquadrant}, does not depend on the assumed expansion history (except in the sense that the assumed expansion history can slightly affect how much of the parameter space can satisfy both this condition and the gradient stability condition).

\begin{figure}[t]
    \centering
    \includegraphics[width=\linewidth]{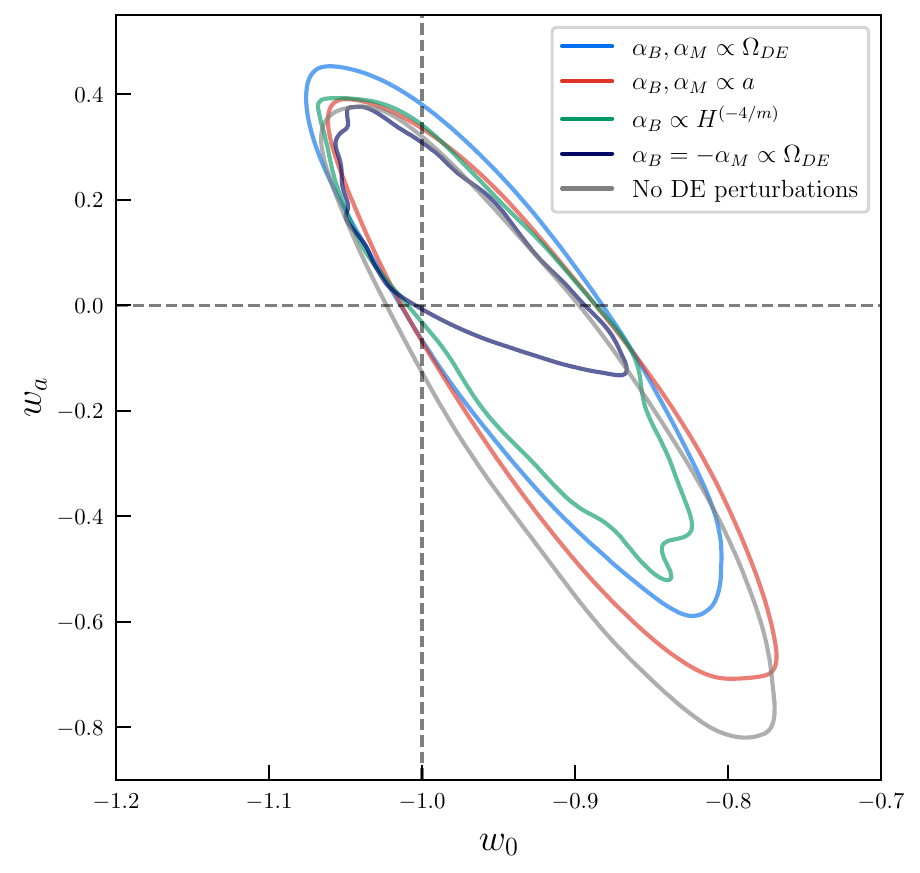}
    \caption{Comparison of constraints on $\{w_0, w_a\}$ in various EFTDE models with the case without any DE perturbations. For clarity, we are only showing the 2$\sigma$ contours. We find that EFTDE models impose a theoretical prior on the background expansion which depends on the set of functional degrees of freedom in the perturbations and their time dependence.}
    \label{w0wa_comparison}
\end{figure}

\begin{figure}[t]
    \centering
    \includegraphics[width=\linewidth]{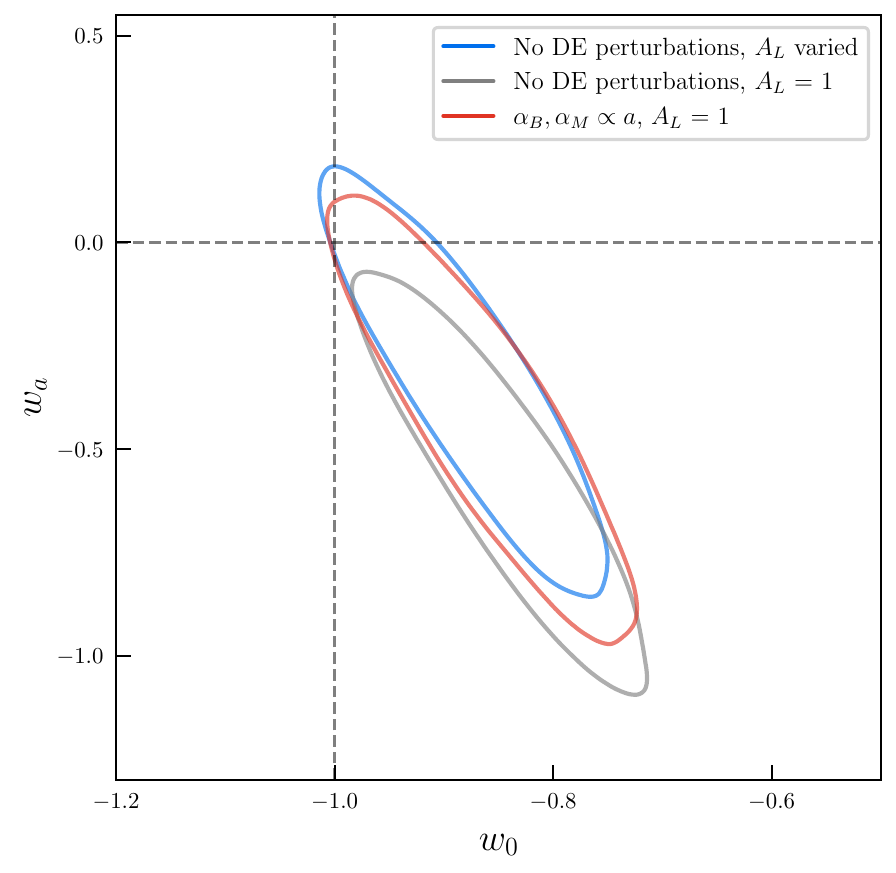}
    \caption{\togglered{Comparison of constraints on $\{w_0, w_a\}$ using data from DESI DR2 BAO, Pantheon+ supernovae and the Planck CMB, between the EFTDE model with $\alpha_B, \alpha_M \propto a$ with the case without any DE perturbations. We fixed $A_L = 1$ for these constraints since this was the choice made for the DESI results \cite{DESI:2025zgx}, and we find that marginalising over $A_L$ has a significant impact on the constraints on $\wzwa$, as we also show in the figure. For clarity, we are only showing the 2$\sigma$ contours. We find that the theoretical prior on $\wzwa$ in the EFTDE model makes the expansion history more consistent with $\Lambda{}$CDM, as it disfavours expansion histories with a large $w_0$ and small $w_a$. This effect of the theoretical prior is more prominent here than in \cref{w0wa_comparison} which shows the constraints using eBOSS BAO+RSD data and marginalising over $A_L$ instead of DESI DR2 BAO and fixing $A_L = 1$, since for the dataset choices in \cref{w0wa_comparison} the constraints on $\wzwa$ are consistent with $\Lambda{}$CDM even for the model with no DE perturbations without any theoretical prior on $\wzwa$.}}
    \label{w0wa_comparison_desi}
\end{figure}

While the effect of freeing up the background expansion history on the constraints on modified gravity perturbations is small in both the phenomenological and EFTDE parametrisations, in EFTDE models the converse is not always true. Through the gradient stability condition, the choice of parametrisation of the EFTDE perturbations places a stronger theoretical prior on the background expansion history than the background parametrisation does on the perturbations. We see this in \cref{w0wa_comparison}, which compares the posterior of $\{w_0,w_a\}$ across the various EFTDE models analysed in the previous sections, along with a purely phenomenological case where DE perturbations are completely neglected and the background expansion history is assumed to be CPL.

\cref{w0wa_comparison} shows that any EFTDE parametrisation of perturbations puts a theoretical prior on $\wzwa$ which affects their observational posteriors. Comparing the background constraints of the $\propto \Omega_{DE}$ and the $\propto a$ time dependences, we see that the more slowly varying time dependence $\propto a$ allows for more freedom in $\wzwa$, in fact it allows almost as much freedom as the purely phenomenological case with no DE perturbations. Further, we find that the assumption of shift symmetry places a strong theoretical prior on $\wzwa$, and even the high generality of the time dependence \cref{shiftsym} is not sufficient to broaden the $\wzwa$ posterior more than the theories that free up $\alpha_M$. Lastly, the strongest theoretical prior on $\wzwa$ is found in the GW background-motivated theory $\alpha_B = -\alpha_M \propto \Omega_{DE}$, which is expected because we see in \cref{omegaDE_GW_w0wa} that in this case the parametrisation of the background expansion history also has a profound impact on the posterior for the perturbations. From \cref{w0wa_comparison}, we can draw a general conclusion that models with only one functional degree of freedom among the $\{\alpha_i\}$, which include the shift symmetric theory and the GW background-motivated theory, impose very strong theoretical priors on $\wzwa$.

\togglered{Recently, BAO observations from DESI \cite{DESI:2024mwx,DESI:2025zgx,DESI:2025fii} have shown tantalising hints for a deviation from a $\Lambda$CDM expansion history in the $\wzwa$ plane. This motivates us to ask if any of the EFTDE models we consider can match the $\wzwa$ values preferred by DESI observations. Specifically, we consider the constraints on $\wzwa$ from the combination of DESI DR2 BAO, CMB measurements from Planck and ACT\footnote{\togglered{Note that our CMB likelihood differs from \cite{DESI:2025zgx} as we don't use ACT measurements and use different Planck likelihoods, but this difference has little impact on the $\wzwa$ constraints, as argued by \cite{ACT:2025tim}.}}, and supernovae from Pantheon+ as obtained in \cite{DESI:2025zgx} (to have our datasets as close as possible to the DESI collaboration's choices, we omitted ISW and RSD data). Given that in \cref{w0wa_comparison}, the $\wzwa$ constraints are least affected by a theoretical prior when the model for the DE perturbations is $\alpha_B, \alpha_M \propto a$, we focus on this model in particular, and check whether the $\wzwa$ constraints are affected by a theoretical prior when the data prefers a background expansion significantly deviating from $\Lambda$CDM.}

\togglered{We compare constraints on $\wzwa$ for the EFTDE model with $\alpha_B, \alpha_M \propto a$ with the constraints on them for the model with no DE perturbations in \cref{w0wa_comparison_desi}. It turns out that the theoretical prior on $\wzwa$ is now strong enough to affect the constraints, and we find that for the $\alpha_B, \alpha_M \propto a$ model, the posterior favours an expansion history more consistent with $\Lambda$CDM than the model with no DE perturbations. This comes from the dependence of the sound speed \cref{cs2} on both $\wzwa$ and $\BMz$, as a result of which the gradient stability condition forces the DE perturbations to deviate strongly from GR if the background deviates strongly from LCDM. While the BAO and supernovae datasets are agnostic to DE perturbations, the CMB isn't, due to the impact of DE perturbations on the late ISW effect and CMB lensing as explained in \cref{sec-data}. Thus, the shift of $\wzwa$ constraints towards LCDM here comes from the joint theoretical prior on $\wzwa$ and $\BMz$ from the gradient stability condition and the constraining power of the CMB on DE perturbations. We also found that the treatment of $A_L$ has an impact on the $\wzwa$ constraints (as found before by \cite{Park:2024pew}), as we show in \cref{w0wa_comparison_desi} by comparing constraints with $A_L$ fixed vs. varied with no DE perturbations. The treatment of $A_L$ also impacts constraints on modified gravity perturbations, a detailed analysis of which is performed in \cref{app-AL}.}

Finding models within EFTDE that reproduce exotic expansion histories such as the one inferred from DESI BAO is challenging, but possible -- see \cite{Ye:2024ywg,Wolf:2024stt,Ishak:2024jhs,Pan:2025psn,Wolf:2025jed} for examples where such models have been found and constrained with data. An alternative approach to study the interplay of the background and perturbations is to start with a Lagragian where the background and perturbation degrees of freedom are not explicitly separated, and map the theory space to the $\{w_0, w_a\}$ parameter space to derive theoretical priors on them, along the lines of \cite{Wolf:2024eph,Wolf:2025jlc}.

\begin{table*}[]
    \centering
\renewcommand{\arraystretch}{1.5}
\begin{tabular}{|c||c|c|c|c|}
\hline
\backslashbox{Model}{Parameter} & $\mu_{\rm{today}}-1$ & $\Sigma_{\rm{today}}-1$ & $\alpha_{B,0}$ & $\alpha_{M,0}$ \\
\hline\hline
$\alpha_B, \alpha_M \propto \Omega_{DE}$ & $0.10^{+{0.05}}_{-{0.10}}$ & $-0.08^{+{0.06}}_{-{0.06}}$ & $0.13^{+0.15}_{-0.23}$ & $0.64^{+0.26}_{-0.48}$ \\
\hline
$\mu - 1, \gamma - 1 \propto \Omega_{DE}$ & $0.12^{+0.19}_{-0.23}$ & $-0.06^{+{0.06}}_{-{0.06}}$ & - & - \\
\hhline{|=||=|=|=|=|}
$\alpha_B, \alpha_M \propto a$ & $0.04^{+{0.05}}_{-{0.07}}$ & $-0.05^{+{0.05}}_{-{0.05}}$ & $0.42^{+0.23}_{-0.39}$ & $0.31^{+0.10}_{-0.26}$ \\
\hline
$\mu - 1, \gamma - 1 \propto a$ & $0.02^{+0.10}_{-0.12}$ & $-0.05^{+{0.06}}_{-{0.06}}$ & - & - \\
\hhline{|=||=|=|=|=|}
$\alpha_B \propto H^{(-4/m)}$ & $0.04^{+0.01}_{-0.04}$ & $=\mu_{\rm{today}}-1$ & $0.70^{+0.45}_{-0.70}$ & - \\
\hline
$\alpha_B \propto H^{(-4/m)}$, $w_0, w_a$ varied & $0.04^{+{0.00}}_{-{0.04}}$ & $=\mu_{\rm{today}}-1$ & $0.43^{-0.07}_{-0.43}$ & - \\
\hline
$\alpha_B \propto \Omega_{DE}$ & $0.03^{+0.01}_{-0.03}$ & $=\mu_{\rm{today}}-1$ & $0.09^{+0.03}_{-0.09}$ & - \\
\hline
$\mu - 1 = \Sigma - 1 \propto \Omega_{DE}$ & $0.03^{+{0.01}}_{-{0.03}}$ & $=\mu_{\rm{today}}-1$ & - & - \\
\hhline{|=||=|=|=|=|}
$\alpha_B, \alpha_M, \alpha_T \propto \Omega_{DE}$ & $0.09^{+0.08}_{-0.10}$ & $-0.08^{+{0.06}}_{-{0.06}}$ & $0.11^{+0.19}_{-0.26}$ & $0.62^{+0.34}_{-0.56}$ \\
\hline
$\alpha_B = -\alpha_M \propto \Omega_{DE}$ & $-0.02^{+{0.02}}_{-{0.01}}$ & $-0.07^{+{0.06}}_{-{0.03}}$ & $-0.14^{+0.12}_{-0.05}$ & $=-\alpha_{B,0}$ \\
\hline
$\alpha_B = -\alpha_M \propto \Omega_{DE}$, $w_0, w_a$ varied & $-0.02^{+0.02}_{-0.04}$ & $-0.07^{+0.05}_{-0.05}$ & $-0.12^{+0.10}_{-0.08}$ & $=-\alpha_{B,0}$ \\
\hhline{|=||=|=|=|=|}
$\alpha_B, \alpha_M \propto \Omega_{DE}$, $w_0, w_a$ varied & $0.11^{+{0.07}}_{-{0.11}}$ & $-0.09^{+{0.06}}_{-{0.07}}$ & $0.30^{+0.17}_{-0.33}$ & $0.82^{+0.31}_{-0.64}$ \\
\hline
$\mu - 1, \gamma - 1 \propto \Omega_{DE}$, $w_0, w_a$ varied & $0.19^{+{0.21}}_{-{0.21}}$ & $-0.07^{+{0.06}}_{-{0.06}}$ & - & - \\
\hline
\end{tabular}

    \caption{Constraints on the parameters describing modified gravity perturbations for all the models discussed in the main text. This table shows the means and the 1$\sigma$ uncertainties on the parameters.}
    \label{tab:parameter_constraints}
\end{table*}

\section{Conclusions} \label{sec-conclusions}
In this paper we have investigated the effects of various theoretical priors on constraints on the nature of dark energy/modified gravity. We especially focused on parametrisations of the dynamics of linear perturbations, which fall into two broad classes: I) Phenomenological $\{\mu,\Sigma\}$ parametrisations that directly model the effect dynamical dark energy has on the Poisson equation(s), and II) Theory-informed effective field theory of dark energy (EFTDE) parametrisations, where the space of functions controlling the dynamics of perturbations is derived from first principles by considering the space of possible interactions given the symmetries of cosmological space-times. Our key findings are as follows:

\begin{itemize}
\item Deriving dark energy dynamics from an underlying EFTDE places a strong theoretical prior on the present-day values of the phenomenological modifications to gravity, $\mSt$. This EFTDE prior both restricts the allowed values that $\mSt$ can take, and introduces correlations between them. In particular, $\mt < 1, \St > 1$ is excluded by the gradient stability condition for the case of a luminal GW speed. Moreover, while the $\mt > 1,  \St < 1$ region in parameter space has a comparatively small prior volume in the EFTDE context, present-day data constraints identify precisely this region as having the largest relative volume in the posterior.

\item A detailed comparison of the predictions following from assuming two different time dependences for the underlying theory functions controlling perturbations -- $\propto \Omega_{DE}$ and $\propto a$ -- revealed a qualitative difference between these two cases in terms of the proximity of the stable and unstable parameter spaces (with respect to gradient stability) in the $\mSt$ space, which is explained by the difference in the properties of the two time dependences near the gradient stability divide, $c_s^2 \simeq 0$.

\item We analysed a rare case of a time dependence that is theoretically motivated in itself, rather than being a phenomenological ansatz -- specifically one derived for a subclass of the EFTDE derived from shift-symmetric scalar-tensor models \cite{Traykova:2021hbr}. Importantly this subclass of theories gives rise to a very specific constraint on the phenomenological modifications to gravity, namely the no-slip condition which is equivalent to $\mu(a) = \Sigma(a)$. We compared various theory-motivated and phenomenological parametrisations that satisfy the no-slip condition, and found that they are not distinguishable from their posteriors on $\mt$.

\item A different class of powerful theoretical priors on the EFTDE comes from gravitational wave physics and we investigated the effect of two such priors: 1) In light of caveats in mapping the observed GW170817 constraint to cosmological scales (an extrapolation of close to 20 orders of magnitude in energy scale) \cite{deRham:2018red}, we investigated the impact of freeing up the cosmological speed of GWs on the constraints on $\mSt$, and found that it leads to a slight broadening of the posteriors -- cf. an analogous analysis and result for EFTDE parameters in \cite{Noller:2020afd}. Importantly, it allows access to $\mSt$ values where $\mt < 1, \St > 1$, which is not possible in models with a luminal GW speed. 2) Requiring the stability of DE perturbations on a GW background sourced by a loud binary black hole merger places additional strong constraints on the EFTDE parameters \cite{Creminelli:2019kjy, Noller:2020afd}. These theories have only one functional degree of freedom to describe DE perturbations, which results in a one-dimensional posterior for $\mSt$ in the context of the time dependences considered here. This leads to significantly tighter constraints than for the two-dimensional case. Moreover, in these theories, the dynamics of the background and perturbations is coupled, and we find profound effects of the background parametrisation on the constraints on perturbations and vice versa.

\item \togglered{Finally, we considered the interplay of the constraints on the background and perturbation dynamics for various classes of EFTDE theories more generally. The gradient stability condition in EFTDE theories couples the background and perturbations. We find that the impact of the choice of model for the background expansion on the constraints on perturbations is relatively small, however the impact of the choice of the model for the perturbations on the constraints on the expansion history is significant. For theories with only one functional degree of freedom for the perturbations -- as discussed in the context of gravitational wave priors and shift symmetry -- the theoretical prior on $\wzwa$ is very strong and prevents significant deviations from $\Lambda{}$CDM in their posteriors. For theories with two functional degrees of freedom $\{\alpha_B, \alpha_M\}$ describing the perturbations, the theoretical prior on $\wzwa$ is weaker, however, these theories are still unable to reproduce the strong deviations from $\Lambda{}$CDM that are seen without any theoretical priors from the combination of DESI DR2 BAO, Pantheon+ supernovae, and Planck CMB data.}

\end{itemize}

The work presented here suggests a number of natural next steps in improving the understanding of theoretical priors and of their effect on observational constraints. The datasets used here, while being very powerful as a result of combining probes of clustering, lensing and cross-correlations, can be complemented by other recent data sets and cosmological probes of gravity. It would be worthwhile to extend our observational analyses to incorporate further recent datasets, such as Planck PR4 \cite{Tristram:2023haj}, ACT DR6 \cite{ACT:2025fju}, and SPT-3G \cite{SPT-3G:2025bzu} for the CMB, and DES \cite{DES:2021wwk,DES:2022ccp} and DESI \cite{DESI:2024mwx,DESI:2025fii,DESI:2025zgx,Ishak:2024jhs} for galaxy clustering. Similarly, it will be intriguing to see what additional constraints the upcoming Euclid mission can place, see e.g. \cite{Euclid:2025tpw}. In a similar vein, it would also be very useful to add a weak lensing dataset to the analysis, in order to further pin down $\St$ and mitigate the degeneracy between $\mSt$ and $A_L$. 

On the theoretical front, there is ample scope to develop further theoretical priors on the EFTDE. This includes I) deriving theoretically motivated time dependences for $\{\alpha_M, \alpha_T\}$ in analogy with \cref{shiftsym} for $\alpha_B$, II) incorporating additional theoretical priors into the analysis, e.g. related to radiative stability \cite{Noller:2018eht,Heisenberg:2020cyi} or from positivity bounds arising as a result of demanding a sensible UV completion -- see \cite{Melville:2019wyy,Kennedy:2020ehn,deRham:2021fpu,Melville:2022ykg,deBoe:2024gpf}  and references therein for how this can impact cosmological parameter constraints, III) imposing complementary theoretical priors derived from considering local solar system constraints and their relation to cosmology \cite{Babichev:2011iz,Burrage:2020jkj,Noller:2020lav}, IV) Further going beyond the linear regime and incorporate further data constraints and theoretical priors derived from non-linear scales, e.g. along the lines discussed in \cite{Cusin:2017mzw,Cusin:2017wjg,Thomas:2020duj,Srinivasan:2021gib,Fiorini:2022srj,Brando:2022gvg,Wright:2022krq,Bose:2022vwi}.

In this paper, we have performed a detailed investigation of how fundamental physics assumptions about dark energy propagate to observational posteriors. Going forward, such an understanding is essential for us to be able to correctly interpret the strong constraints we will obtain from Stage-IV LSS surveys, and use them to acquire further insights into the fundamental physics of dark energy and gravity.

\begingroup
\renewcommand{\addcontentsline}[3]{}

\section*{Acknowledgments}
\vspace{-0.1in}
\noindent 
We thank Seshadri Nadathur for several insightful discussions. 
NS is supported by an UK Science and Technology Facilities Council (STFC) studentship (ST/X508688/1) and funding from the University of Portsmouth and University College London. JN is supported by an STFC Ernest Rutherford Fellowship (ST/S004572/1). KK is supported by STFC grant number ST/W001225/1 and ST/B001175/1.
In deriving the results of this paper, we have used the following codes: CLASS \cite{Blas:2011rf}, getdist \cite{Lewis:2019xzd}, hi\_class \cite{Zumalacarregui:2016pph,Bellini:2019syt}, MGCLASS II \cite{Sakr:2021ylx}, and MontePython \cite{Audren:2012wb,Brinckmann:2018cvx}. We have made use of the SCIAMA HPC cluster at the Institute of Cosmology and Gravitation, University of Portsmouth. For the purpose of open access, the authors have applied a Creative Commons Attribution (CC BY) licence to any Author Accepted Manuscript version arising from this work.
\\

\noindent{\bf Data availability} Supporting research data are available on reasonable request from the authors.

\endgroup

\appendix

\section{$A_L$ and degeneracies} \label{app-AL}

One of our most important probes of DE perturbations at high redshifts is CMB lensing. It is highly sensitive to not only the effects on lensing from modified gravity, but also the nuisance parameter $A_L$ which is an artificial rescaling of the lensing potential power spectrum. Owing to the $A_L$ anomaly mentioned in \cref{sec-data}, there are multiple alternative treatments of $A_L$ found in the literature. In the main text, we follow the most conservative approach of using the legacy Planck temperature and lensing likelihoods, and marginalising over $A_L$ to account for the potential systematic that prefers $A_L \neq 1$ in $\Lambda$CDM. In this appendix, we demonstrate the significant impact of the degeneracy between $A_L$ and $\mSt$. We also highlight two important aspects of this degeneracy that are less appreciated in the literature: that the degeneracy can affect constraints on not only $\St$ but also $\mt$, and that the effect of the degeneracy on $\mSt$ posteriors is highly dependent on the parametrisation.

\begin{figure}[t]
    \centering
    \includegraphics[width=\linewidth]{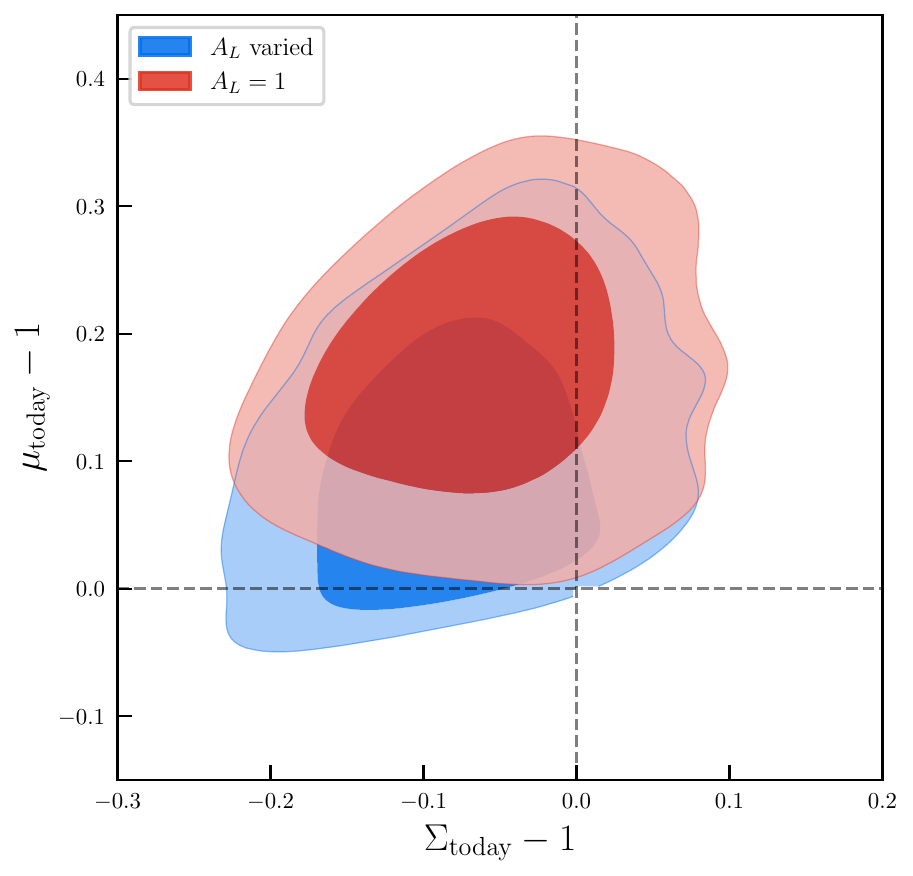}
    \caption{Comparison of constraints on $\mSt$ when $A_L$ is marginalised over with the case where $A_L$ is fixed to $1$ for the $\alpha_B, \alpha_M \propto \Omega_{DE}$ parametrisation. We find that fixing $A_L$ leads to a preference for higher $\mt$, but does not significantly affect the constraints on $\St$.}
    \label{omegaDE_AL_musigma}
\end{figure}

\begin{figure}[t]
    \centering
    \includegraphics[width=\linewidth]{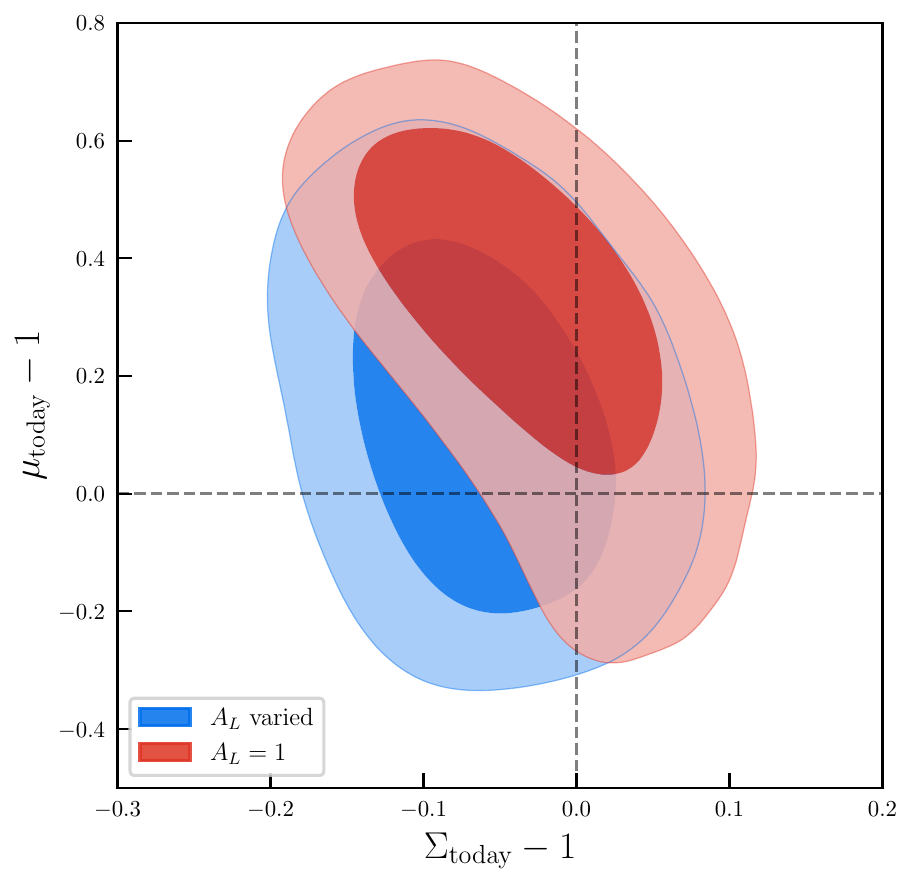}
    \caption{Comparison of constraints on $\mSt$ when $A_L$ is marginalised over with the case where $A_L$ is fixed to $1$ for the $\mu - 1, \gamma - 1 \propto \Omega_{DE}$ parametrisation. We find that fixing $A_L$ leads to a preference for higher $\mt$, but does not significantly affect the constraints on $\St$.}
    \label{Sigmu_AL_comparison_MG}
\end{figure}

First we look at the EFTDE and phenomenological parametrisations with the $\propto \Omega_{DE}$ time dependence, in \cref{omegaDE_AL_musigma,Sigmu_AL_comparison_MG}. In both these cases, fixing $A_L = 1$ results in a posterior that prefers a higher value of $\mt$. Interestingly, the effect on the constraints on $\St$ is smaller than that on $\mt$. This is counterintuitive since it is usually thought that $\Sigma$ is the parameter that affects lensing and $\mu$ affects clustering, and so $A_L$ which affects only the CMB lensing potential is more degenerate with $\Sigma$ than $\mu$. However, it is worth pointing out that according to the Poisson equation \cref{Poisson_Sigma}, the value of the lensing potential depends on both its coupling to the density perturbation, which is controlled by $\Sigma$, and the value of the density perturbation itself, which is affected by any modifications to structure formation, which includes effects from both $\mu$ and $\Sigma$. Thus, it is quite possible for $A_L$ to be degenerate with $\mu$, and the precise effect of the degeneracy on the $\mSt$ posteriors also depends on the constraining power on $\mSt$ of the datasets external to the CMB. In fact, for the EFTDE parametrisation, the constraints on $\mt$ for the $A_L = 1$ case are $\sim 2\sigma$ away from the GR limit (see \cref{tab:parameter_constraints_appendix} for the precise constraints. Note that it also appears from the table that $\mt - 1$ is $\sim 2\sigma$ away from the GR limit for the phenomenological parametrisation, however, from \cref{Sigmu_AL_comparison_MG} we see that the $\mt - 1$ posterior is asymmetric around its mean, thus its "$2\sigma$" uncertainty is not simply twice the $1\sigma$ uncertainty). It would be interesting to see if this preference for $\mt > 1$ persists in the Planck PR4 likelihood \cite{Tristram:2023haj}, and also in the combination of the ACT DR6 \cite{ACT:2025fju} and SPT-3G \cite{SPT-3G:2025bzu} datasets with Planck.

\begin{figure}[t]
    \centering
    \includegraphics[width=\linewidth]{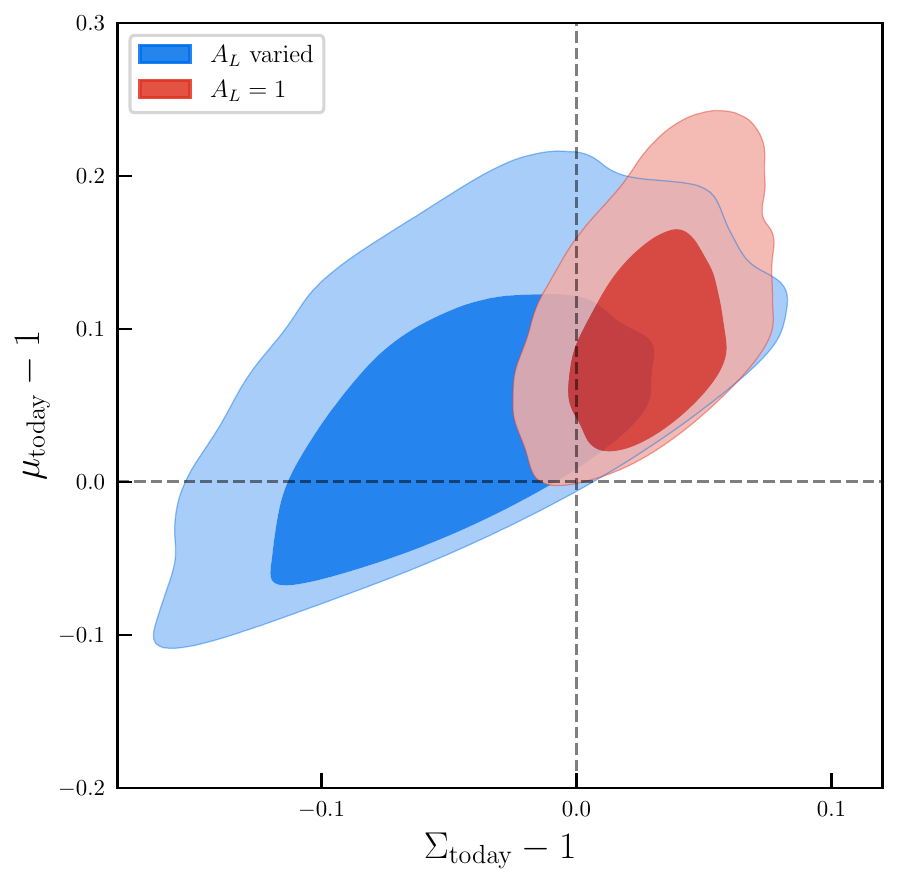}
    \caption{Comparison of constraints on $\mSt$ when $A_L$ is marginalised over with the case where $A_L$ is fixed to $1$ for the $\alpha_B, \alpha_M \propto a$ parametrisation. We find that fixing $A_L$ leads to a preference for higher $\St$, and a milder preference for a higher $\mt$.}
    \label{scale_AL_musigma}
\end{figure}

\begin{figure}[t]
    \centering
    \includegraphics[width=\linewidth]{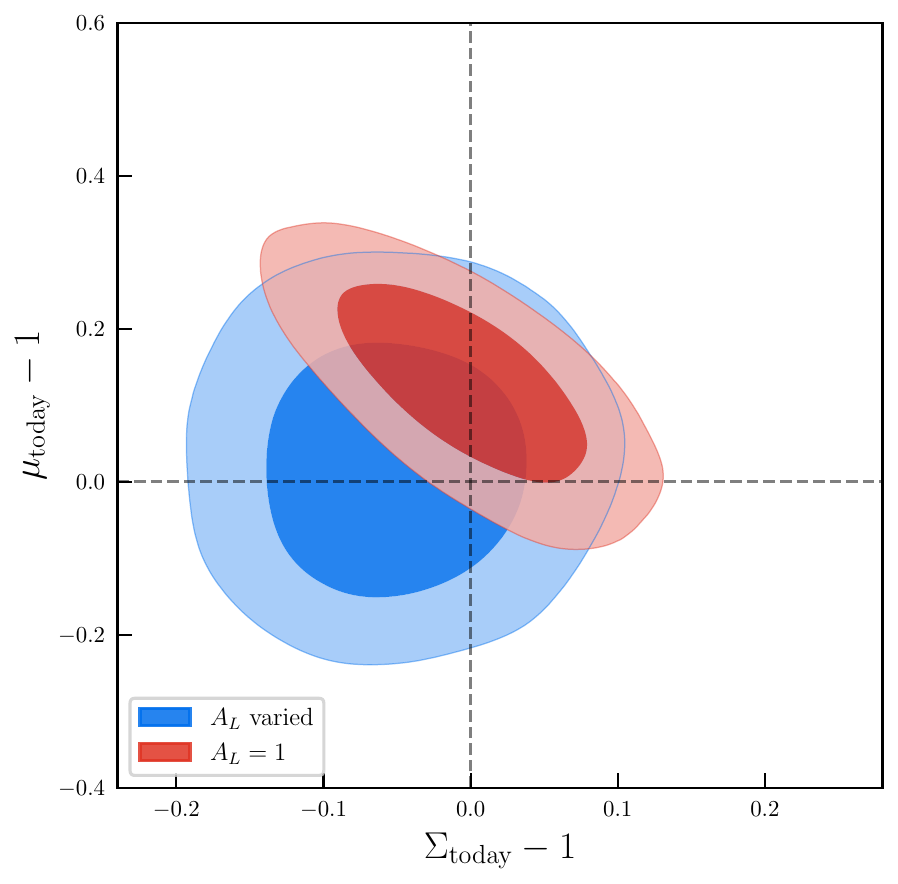}
    \caption{Comparison of constraints on $\mSt$ when $A_L$ is marginalised over with the case where $A_L$ is fixed to $1$ for the $\mu - 1, \gamma - 1 \propto a$ parametrisation. We find that fixing $A_L$ leads to a preference for higher $\mt$, and a milder preference for a higher $\St$.}
    \label{Sigmu_scale_AL_comparison_MG}
\end{figure}

Next, we look at the EFTDE and phenomenological parametrisations with the $\propto a$ time dependence, in \cref{scale_AL_musigma,Sigmu_scale_AL_comparison_MG}. For the EFTDE parametrisation, the effect of fixing $A_L$ looks qualitatively very different for the $\propto a$ time dependence compared to the $\propto \Omega_{DE}$ time dependence. The posteriors for $A_L=1$ prefer a significantly higher $\St$ (and also a less significantly higher $\mt$) compared to the posteriors for varying $A_L$. For both the phenomenological and EFTDE parametrisations, the treatment of $A_L$ affects the shape of the modified gravity posteriors more strongly for the $\propto a$ time dependence than for the $\propto \Omega_{DE}$ time dependence. This is because in the $\propto a$ time dependence, modifications to clustering and lensing extend to higher redshifts (see \cref{musigma_evolution_scale}), and so the CMB lensing likelihood has higher constraining power.

\section{Comparing the $\{\mu, \gamma\}$ and $\{\mu, \Sigma\}$ parametrisations}\label{app-DESode}

\begin{figure}[t]
    \centering
    \includegraphics[width=\linewidth]{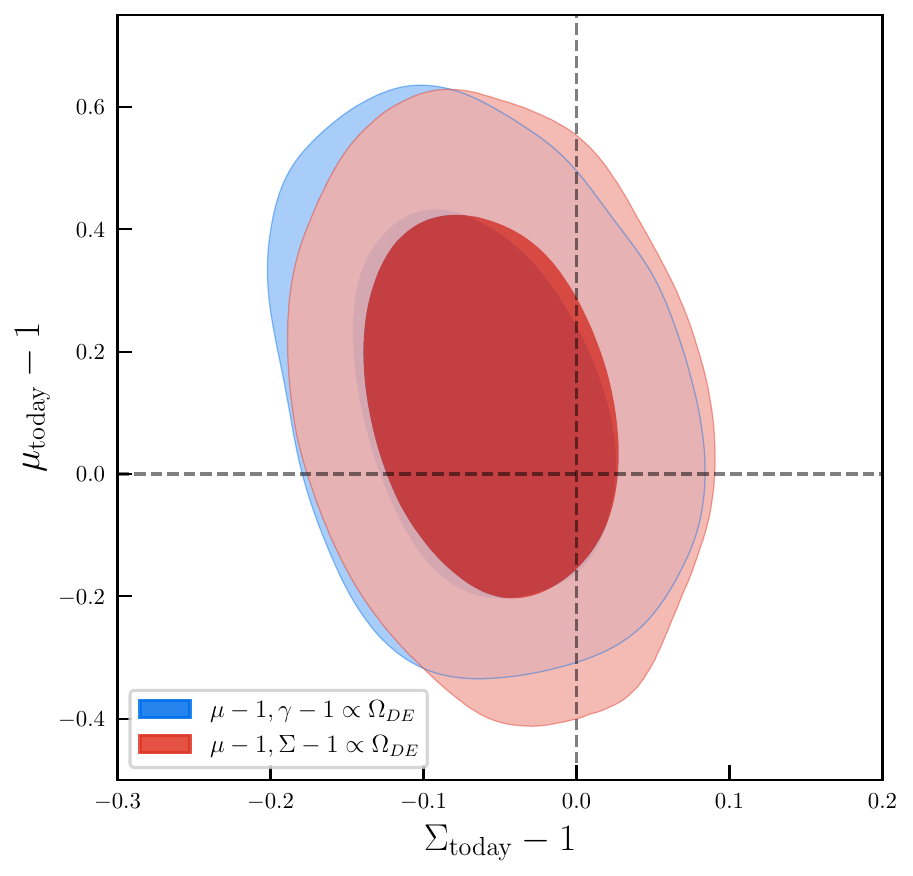}
    \caption{Comparison of constraints on $\mSt$ between the $\mu - 1, \gamma - 1 \propto \Omega_{DE}$ and $\mu - 1, \Sigma - 1 \propto \Omega_{DE}$ parametrisations. We find very similar posteriors for the two parametrisations.}
    \label{Sigmu_DESode_comparison}
\end{figure}

\begin{figure}[t]
    \centering
    \includegraphics[width=\linewidth]{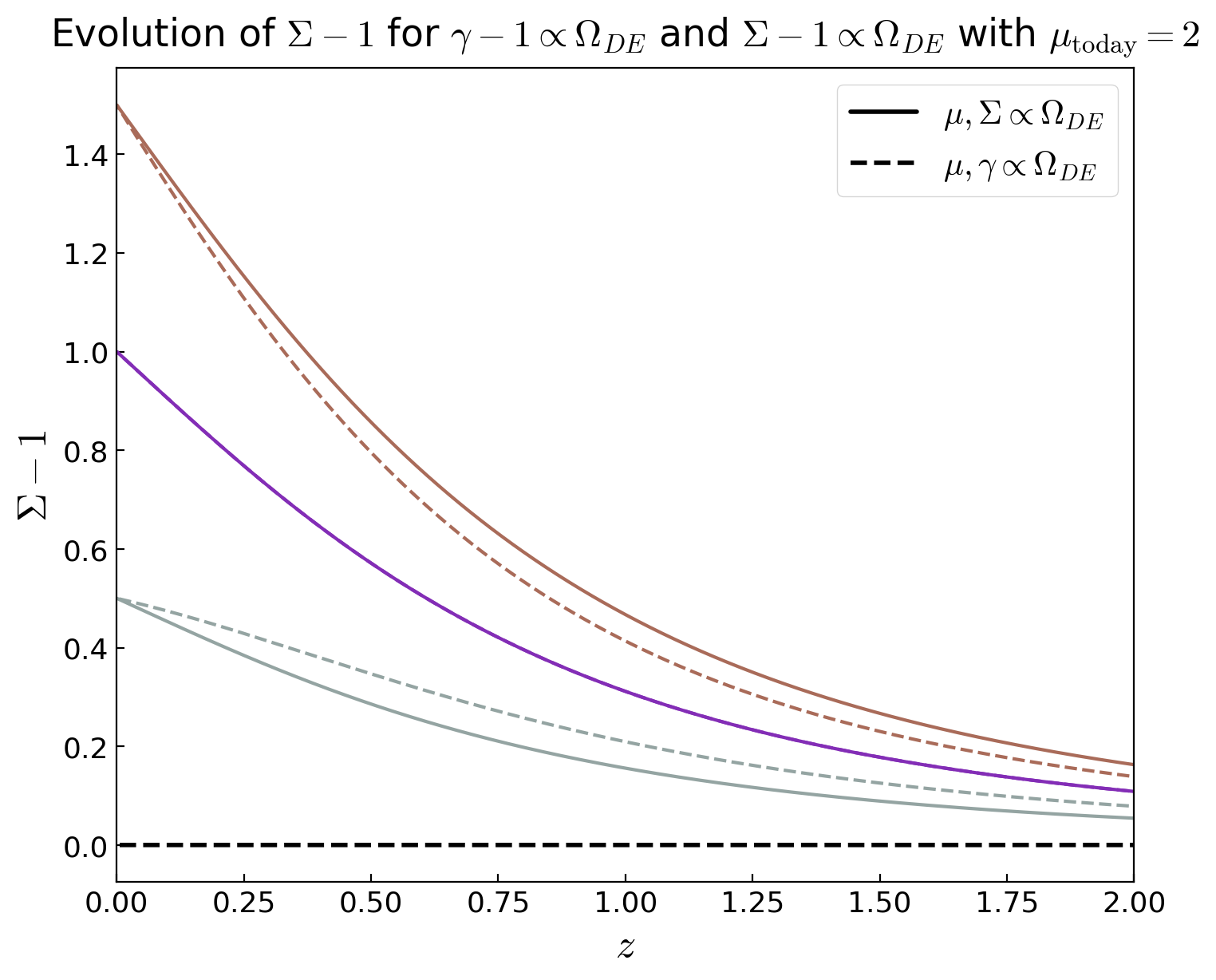}
    \caption{Comparison of the time evolution of $\Sigma(z)$ between the $\mu - 1, \gamma - 1 \propto \Omega_{DE}$ and $\mu - 1, \Sigma - 1 \propto \Omega_{DE}$ parametrisations. We have shown time evolutions for $\mt = 2$ and three different values of $\St - 1$. The solid lines show the time evolution of $\Sigma$ for the $\{\mu, \Sigma\}$ parametrisation, and the dashed lines show the time evolution for the $\{\mu, \gamma\}$ parametrisation.}
    \label{Sigmu_DESode_evolution}
\end{figure}

We find very similar posteriors for the $\{\mu, \gamma\}$ and $\{\mu, \Sigma\}$ parametrisations, as seen in \cref{Sigmu_DESode_comparison}. For the $\mu, \Sigma \propto \Omega_{DE}$ parametrisation, we can write the time dependence of $\gamma$ as a series expansion in powers of $\Omega_{DE}$. Thus, when the present day values $\mSt$ match between the two parametrisations, their time evolution differs at $\mathcal{O}(\Omega_{DE}^2)$, which decays very quickly with redshift. Thus, these two parametrisations give very similar time evolutions for $\{\mu, \Sigma\}$, as can be seen in \cref{Sigmu_DESode_evolution}.

\vfill

\begin{table*}[]
    \centering
\renewcommand{\arraystretch}{1.5}
\begin{tabular}{|c||c|c|c|c|}
\hline
\backslashbox{Model}{Parameter} & $\mu_{\rm{today}}-1$ & $\Sigma_{\rm{today}}-1$ & $\alpha_{B,0}$ & $\alpha_{M,0}$ \\
\hline\hline
$\alpha_B, \alpha_M \propto \Omega_{DE}$, $A_L = 1$ & $0.17^{+{0.07}}_{-{0.07}}$ & $-0.06^{+{0.06}}_{-{0.07}}$ & $0.48^{+0.25}_{-0.29}$ & $1.34^{+0.64}_{-0.73}$ \\
\hline
$\mu - 1, \gamma - 1 \propto \Omega_{DE}$, $A_L = 1$ & $0.31^{+0.24}_{-0.14}$ & $-0.03^{+{0.06}}_{-{0.06}}$ & - & - \\
\hline
$\alpha_B, \alpha_M \propto a$, $A_L = 1$ & $0.10^{+0.03}_{-0.06}$ & $0.03^{+{0.02}}_{-{0.02}}$ & $0.46^{+0.17}_{-0.30}$ & $0.22^{+0.08}_{-0.20}$ \\
\hline
$\mu - 1, \gamma - 1 \propto a$, $A_L = 1$ & $0.13^{+{0.08}}_{-{0.09}}$ & $0.00^{+{0.05}}_{-{0.06}}$ & - & - \\
\hline
$\mu - 1, \Sigma - 1 \propto \Omega_{DE}$ & $0.11^{+{0.21}}_{-{0.20}}$ & $-0.05^{+{0.05}}_{-{0.06}}$ & - & - \\
\hline
\end{tabular}

    \caption{Constraints on the parameters describing modified gravity perturbations for all the models discussed in the appendices. This table shows the means and the 1$\sigma$ uncertainties on the parameters.}
    \label{tab:parameter_constraints_appendix}
\end{table*}

\section{\togglered{The values of $c_s^2$ in the EFTDE parameter space}} \label{app-cs2}

\begin{figure}[t]
    \centering
    \includegraphics[width=\linewidth]{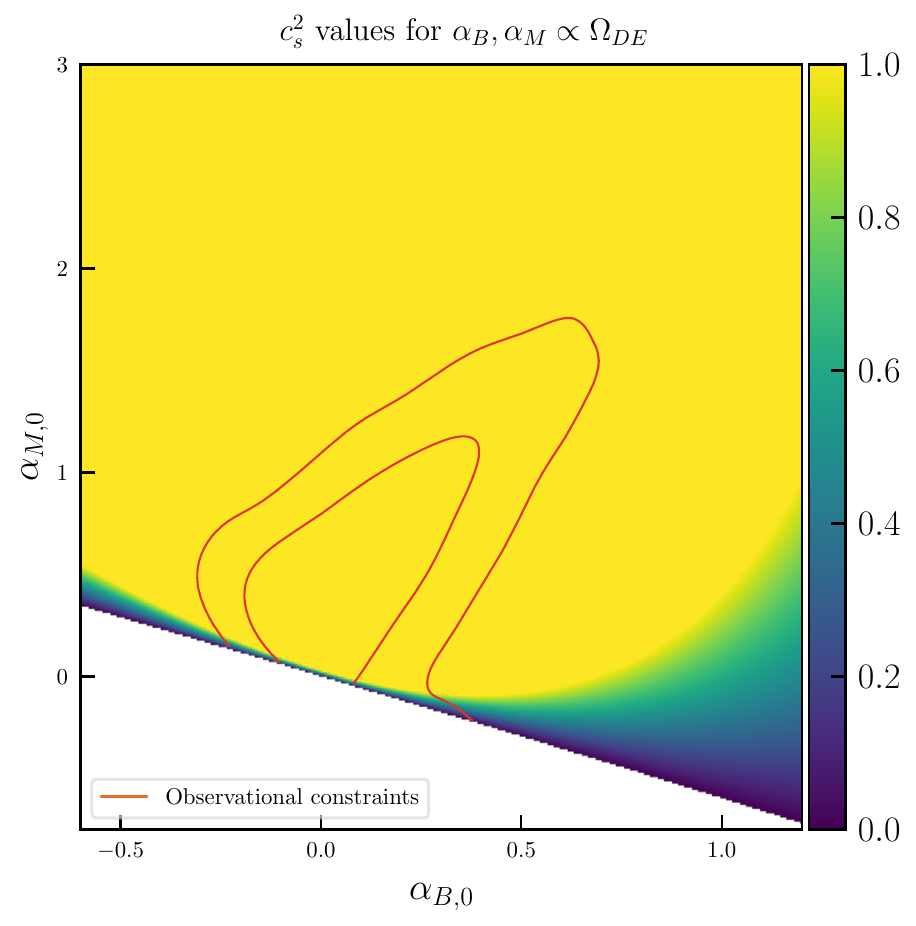}
    \caption{\togglered{The values of $c_s^2$ across the $\BMz$ parameter space for the $\alpha_B, \alpha_M \propto \Omega_{DE}$ model. The range of the colourbar has been limited for clarity, so the yellow region indicates $c_s^2 \geq 1$. It can be seen that most of the parameter space has $c_s^2\ \lsim\ 1$ or $\geq 1$, which is a necessary condition for the QSA as explained in \cref{sec-mapping}. The observational constraints shown in \cref{sec-phenoEFT} are overplotted to show that the QSA is an accurate approximation for almost all of the observationally viable parameter space.}}
    \label{omegaDE_theory_observations_cs2}
\end{figure}

\begin{figure}[H]
    \centering
    \includegraphics[width=\linewidth]{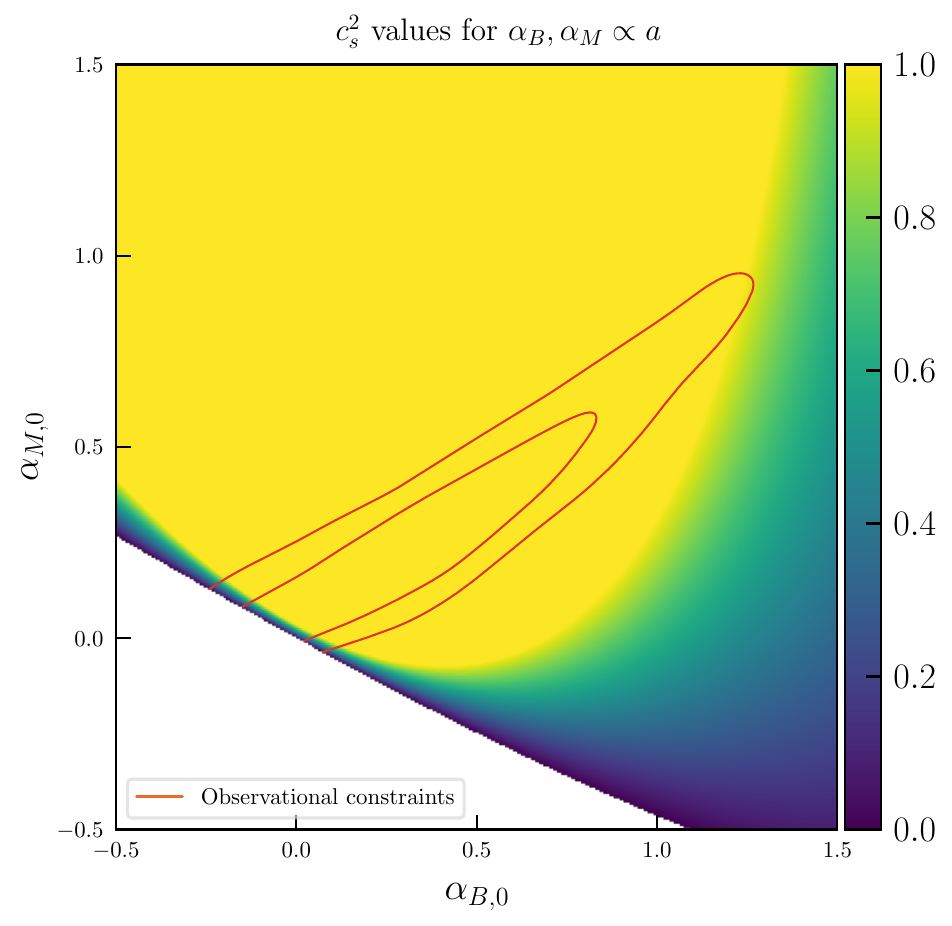}
    \caption{\togglered{The values of $c_s^2$ across the $\BMz$ parameter space for the $\alpha_B, \alpha_M \propto a$ model. The range of the colourbar has been limited for clarity, so the yellow region indicates $c_s^2 \geq 1$. It can be seen that most of the parameter space has $c_s^2\ \lsim\ 1$ or $\geq 1$, which is a necessary condition for the QSA as explained in \cref{sec-mapping}. The observational constraints shown in \cref{sec-scale} are overplotted to show that the QSA is an accurate approximation for almost all of the observationally viable parameter space.}}
    \label{scale_theory_observations_cs2}
\end{figure}

\togglered{Here, we validate the assertion that the sound speed of the scalar field perturbations $c_s^2$ is not $\ll 1$, i.e. is either $\lsim\ 1$ or $\geq 1$ for most of the EFTDE parameter space for two representative models, namely $\alpha_B, \alpha_M \propto \Omega_{DE}$ (\cref{omegaDE_theory_observations_cs2}), and $\alpha_B, \alpha_M \propto a$ (\cref{scale_theory_observations_cs2}). Note that despite the interpretation of $c_s^2 > 1$ indicating a speed faster than light, there is no danger of violating causality, as shown by \cite{Babichev:2007dw,Burrage:2011cr}.}

\vfill\eject

\bibliographystyle{utphys}
\bibliography{TheoryPrior_muSigma_EFT}
\end{document}